\documentclass[journal=nalefd,manuscript=artcile]{achemso}
\usepackage[colorlinks=true,citecolor=blue,urlcolor=blue,linkcolor=blue,bookmarksopen]{hyperref}
\usepackage{siunitx}
\usepackage{chemformula} 
\usepackage{graphicx}
\usepackage{dcolumn}
\usepackage{bm}
\usepackage{amsmath}
\usepackage{amsfonts}
\usepackage{mathtools}
\usepackage{color}
\usepackage{soul}
\usepackage{gensymb}

\renewcommand{\vec}[1]{\boldsymbol{#1}}
\newcommand{\Vpol}{\mathrel{\updownarrow}}
\newcommand{\Hpol}{\mathrel{\leftrightarrow}}
\newcommand{\Dpol}{\mathrel{\text{$\nearrow$\llap{$\swarrow$}}}}
\newcommand{\Apol}
{\mathrel{\text{$\nwarrow$\llap{$\searrow$}}}}

\title{Taming Friedrich-Wintgen interference in resonant metasurface: vortex laser emitting at on-demand tilted-angle}

\author{Raphael Mermet-Lyaudoz}
\affiliation{Univ Lyon, Ecole Centrale de Lyon, INSA Lyon, Universit\'e  Claude Bernard Lyon 1, CPE Lyon, CNRS, INL, UMR5270, Ecully 69130, France}

\author{Cl\'ementine Symond}
\affiliation{Univ Lyon, Universit\'e Claude Bernard Lyon 1, CNRS, Institut Lumi\`ere Mati\`ere, F-69622, Lyon, France}

\author{Florian Berry}
\affiliation{Univ Lyon, Ecole Centrale de Lyon, INSA Lyon, Universit\'e  Claude Bernard Lyon 1, CPE Lyon, CNRS, INL, UMR5270, Ecully 69130, France}

\author{Emmanuel Drouard}
\affiliation{Univ Lyon, Ecole Centrale de Lyon, INSA Lyon, Universit\'e  Claude Bernard Lyon 1, CPE Lyon, CNRS, INL, UMR5270, Ecully 69130, France}

\author{C\'eline Chevalier}
\affiliation{Univ Lyon, Ecole Centrale de Lyon, INSA Lyon, Universit\'e  Claude Bernard Lyon 1, CPE Lyon, CNRS, INL, UMR5270, Ecully 69130, France}

\author{Ga\"{e}lle Tripp\'e-Allard}
\affiliation{Lumi\`ere, Mati\`ere et Interfaces (LuMIn) Laboratory, Universit\'e Paris-Saclay, ENS Paris-Saclay, CNRS, CentraleSup\'elec, 91190 Gif-sur-Yvette, France }

\author{Emmanuelle Deleporte}
\affiliation{Lumi\`ere, Mati\`ere et Interfaces (LuMIn) Laboratory, Universit\'e Paris-Saclay, ENS Paris-Saclay, CNRS, CentraleSup\'elec, 91190 Gif-sur-Yvette, France }

\author{Joel Bellessa}
\affiliation{Univ Lyon, Universit\'e Claude Bernard Lyon 1, CNRS, Institut Lumi\`ere Mati\`ere, F-69622, Lyon, France}

\author{Christian Seassal}
\affiliation{Univ Lyon, Ecole Centrale de Lyon, INSA Lyon, Universit\'e  Claude Bernard Lyon 1, CPE Lyon, CNRS, INL, UMR5270, Ecully 69130, France}

\author{Hai Son Nguyen}
\email{hai-son.nguyen@ec-lyon.fr}
\affiliation{Univ Lyon, Ecole Centrale de Lyon, INSA Lyon, Universit\'e  Claude Bernard Lyon 1, CPE Lyon, CNRS, INL, UMR5270, Ecully 69130, France}
\alsoaffiliation{Institut Universitaire de France (IUF), 75231 Paris, France}

\date{\today}


\begin{document}

	\begin{abstract}Friedrich-Wintgen (FW) interference is an atypical coupling mechanism  that grants loss exchange between leaky resonances in non-Hermitian classical and quantum systems.  Intriguingly, such an mechanism makes it possible for destructive interference scenario in which a radiating wave becomes a bound state in the continuum (BIC) by giving away all of its losses. Here we propose and demonstrate experimentally an original concept to tailor FW-BICs as polarization singularity at on-demand wavevectors in optical metasurface. As a proof-of-concept, using hybrid organic-inorganic halide perovskite as active material, we empower this novel polarization singularity to obtain lasing emission exhibiting both highly directional emission at oblique angles and polarization vortex in momentum space. Our results pave the way to  steerable coherent emission with tailored polarization pattern for applications in optical communication/manipulation in free-space, high-resolution imaging /focusing and data storage.
	\end{abstract}

    \section{Introduction}
	Vector vortex light are optical beams having non-uniform polarization pattern that rotates around a singularity point and exhibiting intensity cross section in form of a doughnut shape\,\cite{Gori:01,Zhan:09,Rosales_Guzm_n_2018}. Over the last years, these exotic light sources have aroused great interests due to tremendous applications in quantum optics\,\cite{Kagalwala2013,PhysRevA.92.023827}, optical manipulation/trapping\,\cite{Kozawa:10,Skelton:13}, light tight-focusing\,\cite{PhysRevLett.91.233901}, optical communication in free-space\cite{Ndagano:15,Ndagano:18} and data storage\,\cite{Parigi2015}. Notably, the generation of vector vortex beams often relies on bulky and/or complex setups: transformation of a Gaussian beam in free space or guided light via optical elements\,\cite{Bomzon:02,Phelan:11,Cai2012,Arbabi2015},  integrated circuit including lasers with integrated polarization or phase modulation components\,\cite{Zheng:10}, combining external frequency feed-back and anisotropic media\,\cite{PhysRevLett.119.113902}. Thus making on-chip vector vortex laser that gets rid of all external components on one hand, and having advantages in terms of robustness, efficiency and reliability on the other, is an appealing research topic of contemporary photonics.  
	
	An elegant strategy to make on-chip vector vortex laser is to use optical bound states in the continuum (BICs) in photonic lattice\,\cite{Hsu2013,Hsu2016,Joseph2021}. They are peculiar photonic resonances forbidden to radiate into free space despite lying in a continuum of propagating waves. As consequence, the radiation pattern of isolated BICs exhibits polarization vortex in momentum space\,\cite{Zhang2018,Doeleman2018} with the vorticity dictated by the topological nature of the resonances\,\cite{Zhen2014,Yoda2020,Ye2020}. When combined with high gain materials, one may obtain lasing action of which the emission is a vector vortex beam pinned at the BIC location in momentum space. Up to now, vector vortex lasers based on BIC have been reported for different lattice geometries and gain materials\,\cite{Iwahashi:11,Iwahashi:11,Kitamura:12,Huang2020,tian2021phasechange,ardizzone2022polariton,Sang2022}. Remarkably, the polarization pattern or the lasing wavelength of BIC vector beams could be dynamically tuned by harnessing non-linearity originated from free-carriers injections\,\cite{Huang2020}, phase change material\,\cite{tian2021phasechange} or exciton-polariton strong coupling regime\,\cite{ardizzone2022polariton}. However, to our knowledge, all of vector vortex microlasers in the literature are pinned at the $\Gamma$ point of the momentum space. Thus they are not ready for many applications such as light tracking, manipulation or steering without external optical elements. 

    Seminal work of Friedrich and Wintgen (FW) showed that BICs can be obtained by harnessing destructive interference mechanism between two leaky resonances\cite{FW85}. In such a paradigm, one resonance gives away all of its losses to the other one and becomes a FW-BIC. Interestingly, FW-BICs usually take place in the vicinity of the avoid crossing point between the two resonances\,\cite{Joseph2021}. Thus tailoring the avoid crossing point in momentum space is an effective route to design FW-BIC at arbitrary oblique wavevector. This promises a salient configuration to make vector vortex microlasers pointing at high-tilted angle. 

	
	
	In this work, we propose a novel proof-of-concept for on-chip vector vortex laser emitting at high oblique angle. The lasing resonance is an original FW-BIC that is tailored by engineering destructive interference out of a Hermitian triple degeneracy of Bloch resonances in subwavelength metasurface. This concept allows isolating the FW-BIC in both momentum space directions. Moreover, the location of this FW-BIC can be designed on-demand by tuning the in-plane anisotropy of the structure.  We demonstrate experimentally lasing action at 550nm from a halide perovskite metasurface. Angle-resolved combined with polarization resolved measurements reveal a coherent vortex emission, pointing at 21 degrees. All experimental results are perfectly explained by an analytical model based on FW interference mechanism, and nicely matched with numerical simulations. Our demonstrated vector vortex on-chip would open the way for steerable coherent emission with tailored polarization patterns. 
	
	
	\section{Results and discussion}
	
	\subsection{Nanophotonic concept and design} The photonic resonance of our vector vortex laser is designed to exhibit farfield polarization singularity that is located at a photonic band edge of the energy-momentum dispersion $E(k_x,k_y)$. These singularities, corresponding to FW-BICs, is tailored at arbitrary point of the momentum space by engineering FW hybridization between Bloch resonances from subwavelength metasurface. Indeed, a suppression of radiative losses, accompanied by an avoid crossing, results from the interplay between different coupling mechanisms (direct couplings via nearfield effects and radiative couplings via the radiative continuum) in the vincinity of the crossing point for uncoupled resonances when sweeping parameters such as wavevectors, geometrical sizes, refractive index... Here the set of Bloch resonances is chosen so that the crossing point takes places in momentum space with both  $k_x$ and $k_y$ wavevectors as sweeping parameters. Thus the  destructive interference configuration leads to an isolated FW-BIC. This isolated FW-BIC corresponds to a farfield radiation-singularity that is pinned at a photonic band edge with infinite quality factor.     
	
	\begin{figure}[htb!]
		\begin{center}
			\includegraphics[width=16cm]{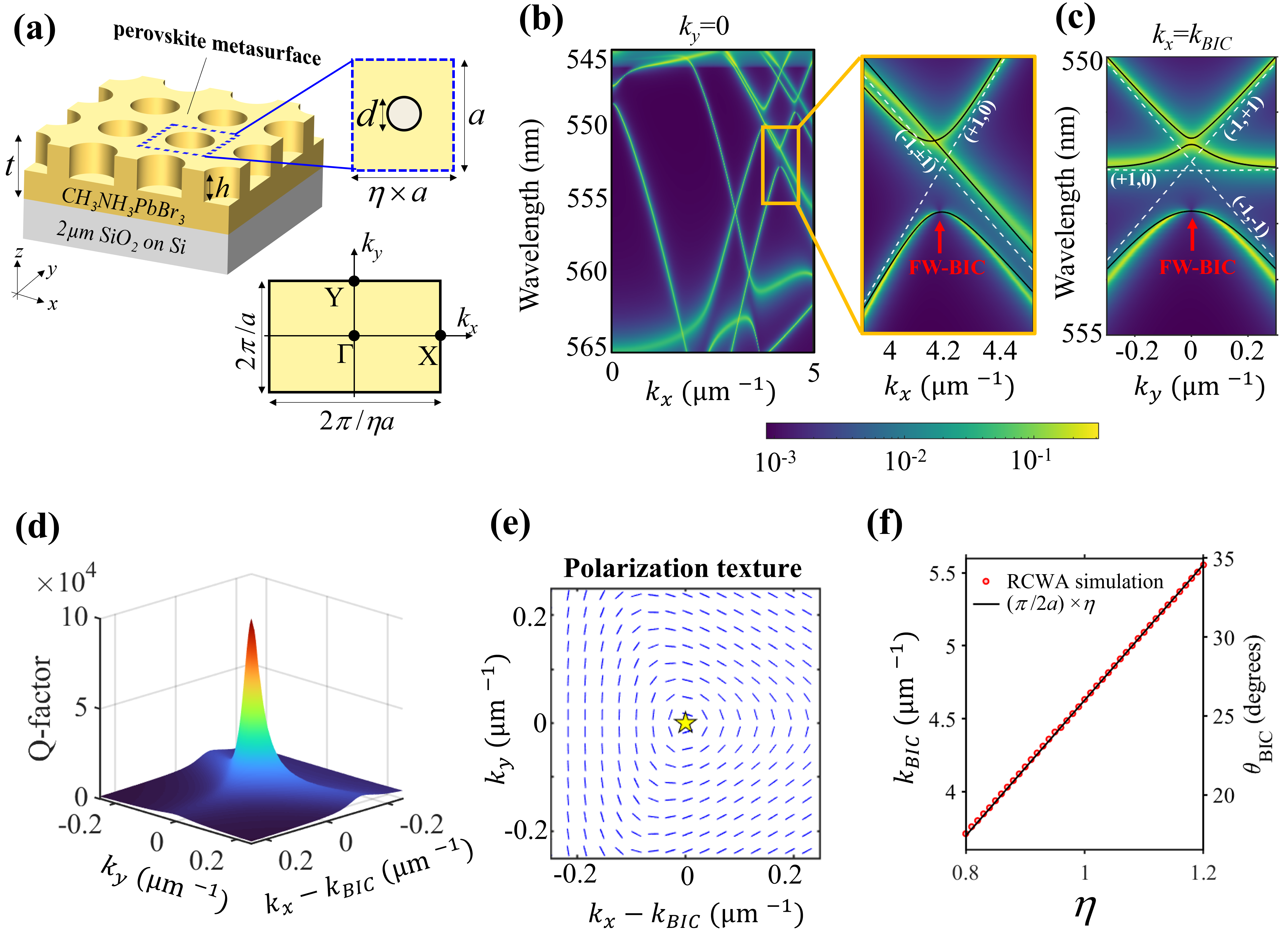}
		\end{center}
		\caption{\label{fig1}{(a) Scheme of the perovskite based metasurface and the corresponding unit cell, and Brillouin zone. The unit cell is a rectangular lattice having $\eta$ as the anisotropy factor of period mismatch along $x$ and $y$-axis. The Brillouin zone is also a rectangular cell having $2\pi/\eta a$ and $2\pi/a$ as size along $k_x$ and $k_y$ respectively.  (b) Angle-resolved absorption simulations along $k_x$ direction, when $k_y=0$. The orange rectangular corresponds to a zoom in to highlight the FW-BIC. (c) Zoom in of the angle-resolved absorption simulation along $k_y$ direction, when $k_x=k_{BIC}$, to highlight the FW-BIC. In both (b) and (c), the FW-BIC is indicated by the red arrow; black lines are fittings using effective theory of FW couplings between the three guided resonances $(-1,\pm 1)$ and $(+1,0)$; white dashed lines indicate the uncoupled modes. (d,e) Results from numerical simulations of the quality factor (d) and polarization texture (e) in the vicinity of the FW-BIC. (f) Location of the FW-BIC in $k_x$ direction as a function of the anisotropy factor $\eta$. Red circles are results from numerical simulations. Black line is the analytical prediction.}}
		\label{design}
	\end{figure}
	
	Hybrid organic-inorganic halide perovskite has been chosen as active material for our proof-of-concept due to their maturity and excellent properties for solution-processable lasers (direct bandgap, high gain, low trap density)\,\cite{Wei2019,Zhang2021}, with lasing effect in perovskite metasurface previously reported by different groups\,\cite{Lei2021,Pourdavoud2017,Li2018,Bar2018,Sun2020,Huang2020,tian2021phasechange}. Our photonic metasurface, depicted in Fig.~\ref{fig1}a, is made of a CH$_3$NH$_3$PbBr$_3$ perovskite slab of thickness $t=270$\,nm. The perovskite slab is partially corrugated by a rectangular lattice of air holes, with patterning depth $h=80$\,nm and diameter $d=200$\,nm. The lattice period along $y$-axis of $a=340$\,nm and the one along $x$-axis of $\eta\times a$ are slightly mismatched, with the anisotropy factor $\eta=0.9$. The anisotropy of the unit cell in real space correspond to the same anisotropy of the Brillouin zone in momentum space, having a $C_2$ symmetry $\Gamma$, X and Y as high symmetry points (see Fig.~\ref{fig1}a). This anisotropy plays a key role to tune the farfield location of the targeted FW-BIC, as explained in the following.
	
	Figure.~\ref{fig1}b shows the simulated photonic band structure of the metasurface along $\Gamma$X direction (i.e. sweeping $k_x$, with $k_y=0$) obtained by absorption calculation using the Rigorous Coupled Wave Analysis (RCWA) (see Methods section). We pay particular attention to three photonic bands shown in the zoom of the simulation results (orange rectangle in Fig.~\ref{fig1}b). These modes are results of the hybridization of three Bloch resonances $(+1,0)$, $(-1,+1)$ and $(-1,-1)$ that are all originated from the fundamental transverse electric guided mode of the perovskite slab. Here the indexes $n$ and $m$ in $(n,m)$ are diffraction orders along $x$ and $y$ direction, corresponding to the propagation vector $\vec{\beta}_{n,m}(k_x,k_y)=\left(\frac{2n\pi}{\eta a}+k_x\right)\vec{u_x} + \left(\frac{2m\pi}{a}+k_y\right)\vec{u_y}$ of the original guided mode. We note that $(-1,+1)$ and $(-1,-1)$ are always degenerated along $\Gamma X$ direction ($k_y=0$),  a triple degeneracy point occurs at $k_x=\frac{\pi}{2a}\eta$ and $k_y=0$. Most importantly,  the coupling between these three  resonances gives rise to an avoid crossing effect that is accompanied by the formation of a FW-BIC at the lowest band (see Methods and the Supplemental Information for details of our analytical theory). Indeed, in the zoom of  Fig.~\ref{fig1}b, we distinguish clearly an avoid crossing of the resulted resonances, and a local vanishing of the lowest resonance at its band-edge ($k_x=k_{BIC}$=\SI{4.16}{\micro\metre}$^{-1}$) that is the hall-mark of a FW-BIC\,\cite{Joseph2021}. Figure.~\ref{fig1}c depicts simulated photonic band structures along $k_y$ when fixing $k_x=k_{BIC}$ in the same spectral range as the zoom in Fig.~\ref{fig1}b, showing the three hybridized  resonances previously discussed.  We distinguish again a local vanishing of the lowest resonance at its band-edge, here at $k_y=0$. As consequence, our photonic structure exhibits a FW-BIC that is localized at $k_x=k_{BIC}$ and $k_y=0$. From Figs.~\ref{fig1}b,c, we note also an excellent agreement between the  band structure obtained from numerical simulations and the theoretical one obtained by our analytical model based on FW theory.
	
	The quality factor (Q-factor) of the lowest band when scanning $k_x$ and $k_y$ is numerically obtained from RCWA simulation, and reported in Fig.~\ref{fig1}d. A Q-factor enhancement at the FW-BIC is clearly evidenced, with $Q\approx 10^5$ at $k_x=k_{BIC}$ and $k_y=0$. A finite numerical value of the Q-factor, unlike an infinite value predicted by the analytical model detailed in the Supplementary Information, can be explained by two reasons: i) parasitic absorption of the perovskite material bellow its band-gap; ii)  imperfect destructive interference of radiative losses due to the presence of other guided mode resonances that are not taken into account in the analytical model. Thus our FW-BIC is of quasi-BIC nature, which may still possess some radiative and non-radiative losses. However, these losses are negligible for lasing applications based-on halide perovskites that only require quality factors in order of $10^3$.\cite{Lei2021}
	
	Figure~\ref{fig1}e presents the polarization texture of the lowest hybrid band, extracted from RCWA calculations. A vortex texture of linear polarization light rotating around the FW-BIC is clearly evidenced. This vortex is associated to a topological charge\,\cite{Zhen2014} of $q=\frac{1}{2\pi} \oint_\mathcal{C} d\vec{k} \nabla_{\vec{k}} \phi=-1$ where $\phi(\vec{k})$ is the orientation of the farfield polarization vector.   We note that the same topological charge exists at $(k_x=-k_{BIC},k_y=0)$ and at the same energy due to the $C_2$ symmetry of our structure. This twin singularity is a FW-BIC resulting from the coupling between $(-1,0)$, $(+1,+1)$ and $(+1,-1)$ guided resonances. Therefore our design suggests an original engineering of slow-light singularities at arbitrary oblique angle.
	
	The location of the FW-BIC in the momentum space can be tailored on-demand via the anisotropy factor $\eta$. Indeed, as shown in Fig.~\ref{fig1}b, $k_{BIC}$ almost coincides to the momentum of the triple degeneracy point between $(+1,0)$ and $(-1,\pm1)$. Therefore the FW-BIC position is given by:
     \begin{equation}
     k_{BIC}\approx \frac{\pi}{2a}\eta.\label{eq:kBIC}
     \end{equation}
     Such a simple dependence is nicely confirmed when compared to the value of $k_{BIC}$ extracted from RCWA simulations for different values of $\eta$ (see Fig.~\ref{fig1}f). In the following, for the experimental demonstration, we chose $\eta=0.9$, so that the FW-BIC corresponds to an angle of $\theta_{BIC}=21.3^o$ which is a high-tilted angle but still inside the numerical aperture of our experimental setup ($\theta_{max}=26.7^o$)..

	\subsection{Experimental results:}
\begin{figure}[htb!]
		\begin{center}
			\includegraphics[width=16cm]{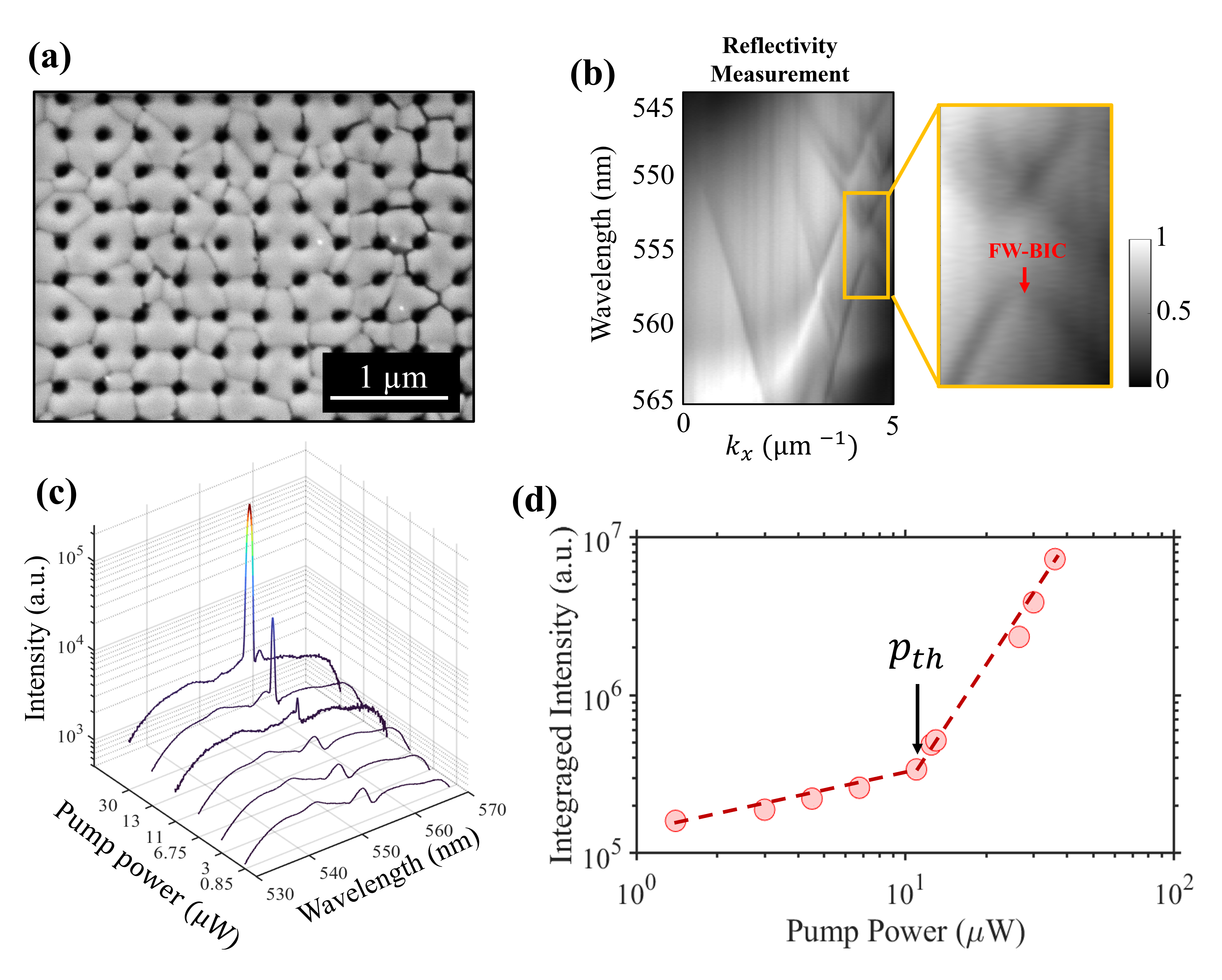}
		\end{center}
		\caption{\label{fig2}{(a) SEM top view of the perovskite metasurface. (b) Angle-resolved reflectivity measurement along $k_x$ and a zoom in the vicinity of the FW-BIC that is indicated by the red arrow. (c) Angle-integrated spectra of photoluminescence at different pump powers. (d) Spectral and angle-integrated intensity as the function of the pump power. The lasing threshold $p_{th}$ is indicated by the black arrow.}}
		\label{model}
	\end{figure}

	The designed metasurface is fabricated by direct thermal nano-imprint lithography on solution-processed CH$_3$NH$_3$PbBr$_3$ perovskite thin film (see fabrication details in Methods). The perovskite layer is first deposited via spincoating onto a substrate of \SI{2}{\micro\metre} of SiO$_2$ on silicon. It is then followed by a nano-imprinting process using a patterned silicon mold.  Figure.~\ref{fig2}a depicts Scanning Electron Microscopy (SEM) image of the fabricated structure.
		
	We first characterize the photonic modes of the perovskite metasurface via angle-resolved reflectivity measurements obtained from a house-built setup. Figure.~\ref{fig2}b depicts the reflectivity spectra along $\Gamma$X direction. The dispersion of Bloch resonances shown in these spectra is in good agreement with the one predicted by the theory and numerical simulations in Fig.~\ref{fig1}b. In particular, the avoid crossings of three hybrid modes, and the formation of FW-BIC in the lower branch are clearly evidenced.  
	
	To investigate lasing action of the fabricated sample, the perovskite metasurface is optically pumped (100\,Hz, 5\,ns pulsed laser at 450\,nm, spot-size diameter of \SI{50}{\micro\metre}). Since the lifetime of polycristalline perovskite is about few ns\cite{Droseros2018}, only a fraction of the 5-ns pulse laser is used for carrier injection while most of the pulse energy is attributed for heating effect. Thus we adopt the average power instead of energy per pulse for evaluating the pump intensity.  The angle-integrated photoluminescence spectra, and spectrally integrated photoluminescence intensity for different pump powers are shown in Fig.~\ref{fig2}c and Fig.~\ref{fig2}d respectively. These results show a lasing action at the threshold $p_{th}\approx \SI{11}{\micro\watt}$: i) Below threshold, the photoluminescence spectra corresponds to a broad spontaneous emission spanning over 40 nm (from 530 to 570 nm). ii) At threshold, a sharp peak of 0.49nm linewidth corresponding to stimulated emission starts emerging from the spontaneous emission background. The gain and loss are balanced, thus we can use the measured linewidth to estimate the quality factor of the lasing mode that amounts to 1130. iii) Above threshold, the emission intensity exhibits a jump and increases drastically. The emission spectrum is completely dominated by the stimulated emission.     
	
	\begin{figure}[htb!]
		\begin{center}
			\includegraphics[width=16cm]{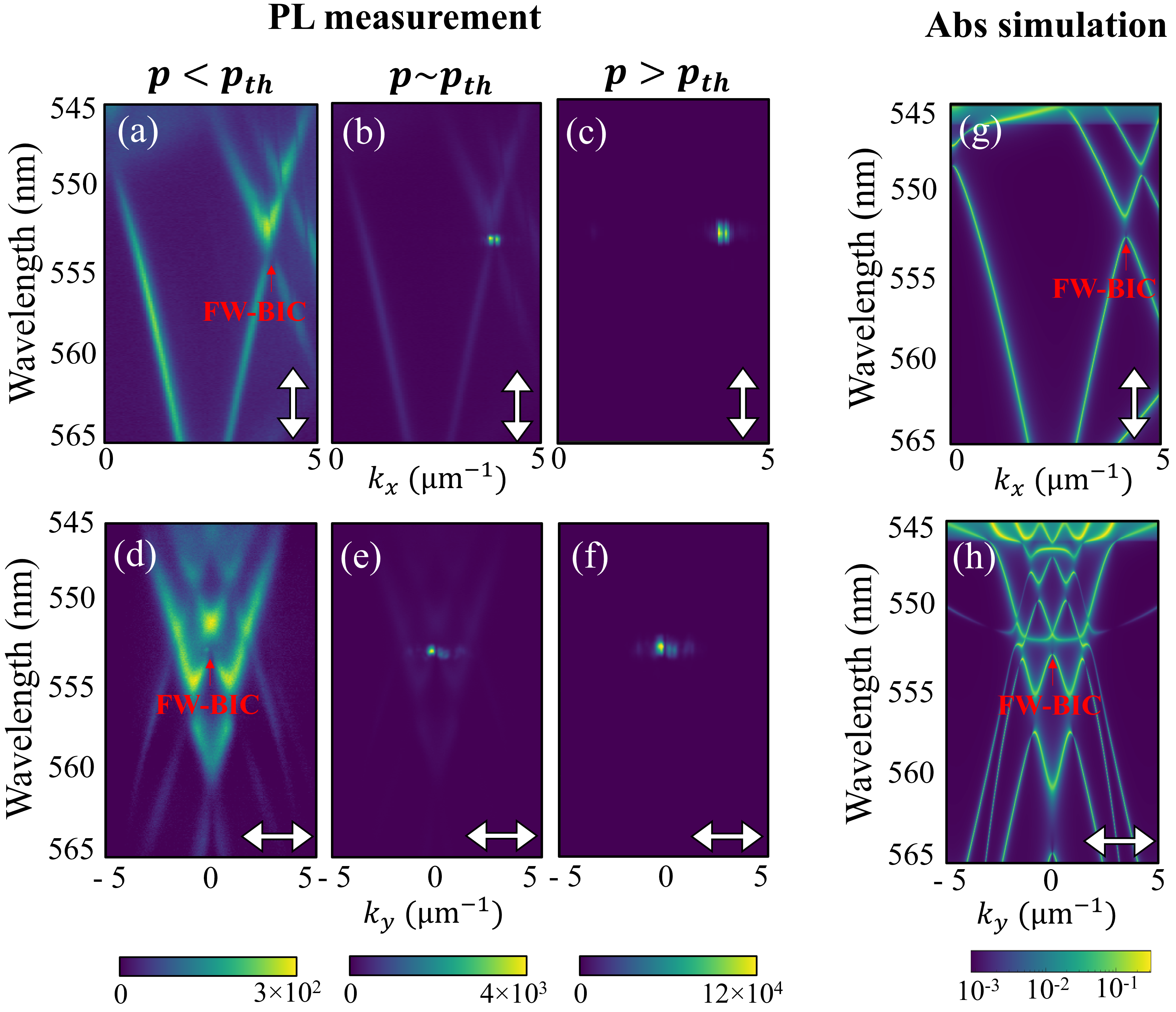}
		\end{center}
		\caption{\label{fig3}{(a-f) Angle-resolved photoluminescence spectra along $k_x$ direction at $k_y=0$ (a-c), and along $k_y$ direction at $k_x=k_{BIC}$ (d-f) for pump power below threshold (a,d), at threshold (b,e) and above threshold (c,f). The emission is filtered along $\Vpol$ for measurements along $k_x$ and $\Hpol$ for measurements along $k_y$. (g,h) Angle-resolved absorption spectra along $k_x$ direcion at $k_y=0$ (g), and along $k_y$ direction at $k_x=k_{BIC}$ (h). }}
		\label{power_increase}
	\end{figure}
	
	The origin of this lasing action is revealed by monitoring the angle-resolved photoluminescence spectra when increasing the pump power across the lasing threshold. Figure.~\ref{fig3} depicts the experimental results of these measurements below threshold [Figs.~\ref{fig3}a,d], at threshold  [Figs.~\ref{fig3}b,e] and above threshold [Figs.~\ref{fig3}c,f] for angle-resolved emission along $k_x$ with $k_y=0$  [Figs.~\ref{fig3}a-c], and along $k_y$ with $k_x=k_{BIC}$  [Figs.~\ref{fig3}d-f].For the sake of clarity, here we only analyse the emission in the polarization of the FW-BIC ($\Vpol$ for $k_x$ dispersions and $\Hpol$ for $k_y$ dispersions). As references, the corresponding angle-resolved absorption spectra are also simulated and shown in Figs.~\ref{fig3}g,h. The FW-BIC is clearly evidenced as the lasing mode: i) Below threshold, the spontaneous emission populates the whole photonic dispersion in our spectral window [Figs.~\ref{fig3}a,d] and these emission spectra are nicely matched  with the simulated absorption ones [Figs.~\ref{fig3}g,h]. ii) At threshold, most of the emission is located spectrally and momentumly at the FW-BIC [Figs.~\ref{fig3}b,e]. This localized signal is the stimulated emission that is nurtured by long lifetime photons of FW-BIC. However, one may still distinguish a weak spontaneous emission background from other photonic modes. iii) Above threshold, only emission in the vicinity of the FW-BIC mode is observed.  Indeed, similar to lasing action at $\Gamma$ point of symmetry-protected BIC, our laser beam in  momentum space exhibits a doughnut shape centred at the BIC\,\cite{Iwahashi:11,Vo:10}. This doughnut shape is projected as two lobes on both sides of the BIC in the energy-momentum dispersion measurements. The doughnut centre imposes the pointing angle and the doughnut diameter (i.e. distance between the two lasing lobes in energy-momentum dispersion measurements) dictates the beam divergence. This corresponds to a directional emission pointing at $k_{BIC}$ and localized in the reciprocal space within $\Delta k_x=$\SI{0.37}{\micro\metre}$^{-1}$. Light emission thus occurs at 21-degree oblique angle  with a divergence less that 2 degrees. Similar lasing action also takes place at the twin FW-BIC and corresponds to a lasing emission at -21 degrees.
	
\begin{figure}[htb]
		\begin{center}
			\includegraphics[width=16cm]{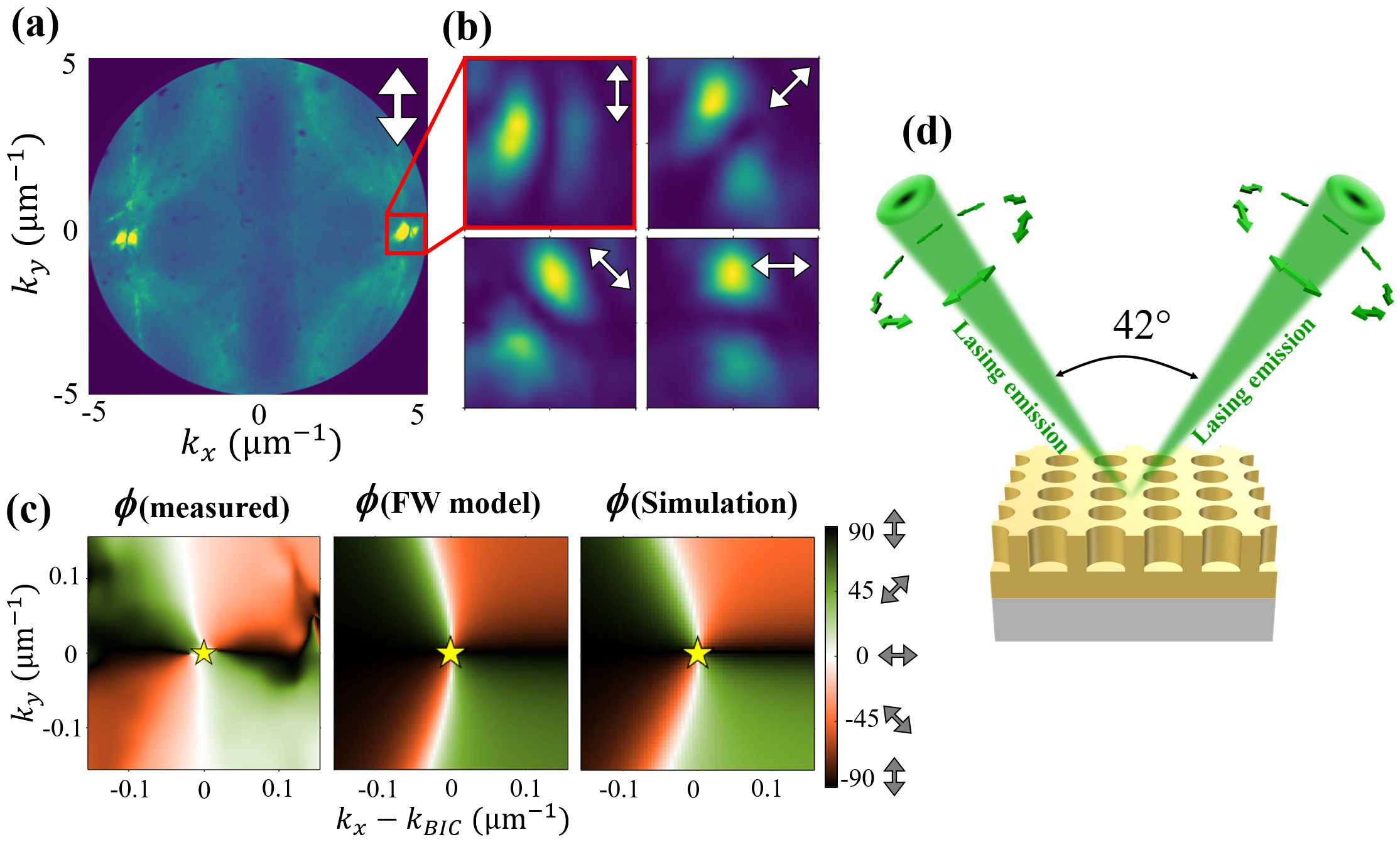}
		\end{center}
		\caption{\label{fig4}{(a) Farfield image of the lasing emission, analysed in $\Vpol$ polarization. (b) Zoom in of the farfield emission in the vicinity of the FW-BIC in four different polarizations $\Vpol$, $\Dpol$, $\Hpol$, and $\Apol$. (c) Polarization texture of the lasing emission extracted from the experimental measurements (left panel), calculated by the FW theory (middle panel), and provided by RCWA simulations (right panel). (d) Illustration of the vortex lasing emissions demonstrated in this work. }}
		\label{disp}
	\end{figure}

	Finally we study the polarization properties of the lasing emission. Its farfield is analysed in four different polarizations $\Hpol$, $\Vpol$, $\Dpol$ and $\Apol$, then is directly imaged  onto the camera sensor without any spectral selection [see Fig.~\ref{fig4}a,b)]. It is shown that only two lobes of the doughnut beam are selected for each polarization. To reconstruct the polarization texture, the Stokes parameters $S_1$ and $S_2$ are extracted from the four intensity mappings: $S_1 =\left(I_{\Hpol}-I_{\Vpol}\right)/\left(I_{\Hpol}+I_{\Vpol}\right)$ and $S_2=\left(I_{\Dpol}-I_{\Apol}\right)/\left(I_{\Dpol}+I_{\Apol}\right)$. Then the polarization orientation angle $\phi$ is given by $\tan{2\phi}=S_1/S_2$. The experimental results of polarization texture are depicted in Fig.\,\ref{fig4}c (left panel), and are in perfect agreement with the analytical results from the effective theory [Fig.\,\ref{fig4}c (middle panel)], and numerical results from RCWA simulations [Fig.\,\ref{fig4}c (right panel)]. It shows that the polarization angle rotates a full 2$\pi$ clock-wise when encircling the singularity point pinned at the FW-BIC. Therefore our lasing emission is a vector vortex light in momentum space with topological charge $q=-1$.  The same topological charge is also observed for the FW-BIC located at $k_x=-k_{BIC}$ (see supplemental material). Therefore our device exhibit two oblique vector vortex lasing emissions as illustrated in Fig.\,\ref{fig4}d.  We note that while high-angle emitting micro-laser using BIC concepts have been recently demonstrated\cite{Ha2018,Sang2022}, the reported lasing actions do not take place at singularity point and exhibits only gaussian beam. Our observation is then the first demonstration of oblique vector vortex lasing action.

	\section{Conclusion}
    	In conclusion, we have proposed and demonstrated experimentally for the first time a micro-laser emitting vector vortex beam at high tilted-angle. The photonic concept is based on an original engineering of FW interference between three guided mode resonances, leading to the formation of polarization singularity at off-$\Gamma$ band-edge. The lasing action is achieved thanks to the implementation of high gain material (i.e. halide perovskite) at the FW-BIC. Moreover, our work provides a unique scheme to tailor vortex beam at arbitrary oblique angle via the use of anistrotropic lattice. Since our proof-of-concept employs perovskite metasurface fabricated by solution-based method and nano-imprint structuration, a beam steering mechanisms can be straightforwardly implemented via the use of flexible or piezoelectric substrates. Therefore, our work would pave the way for dynamical control of vortex beam lasing for applications in optical manipulations, trapping or communication in free space.
	
    \section{Methods}
        \paragraph{Mold fabrication:} A 100 nm-thick hydrogen silsesquioxane (HSQ) resist is spun on a silicon substrate, then baked at 80 $^{\circ}$C for 4 minutes. The resist is exposed by electron-beam lithography to define negative patterns (i.e. pillar lattices) and developed with a solution of TMAH. The patterns are subsequently transferred to the Si substrate by inductively-coupled reactive ion etching (ICP-RIE) using a mixture of Cl$_2$ and O$_2$. 
        
         \paragraph{Perovskite preparation and deposition:} The perovskite solution is prepared by mixing CH$_3$NH$_3$PbBr and PbBr$_2$ (1.2: 1) in N,N-dimethylformamide (DMF), then stirred for 1 h at room temperature. Then \SI{80}{\micro\litre} of the precursor solution is spin-coated at 5000 rpm for 30 s. In the aim to obtain homogeneous and uniform active layer \SI{60}{\micro\litre} of Chlorobenzene antisolvent is dropped at 5s after launching  the spin coating. The sample is then annealed by baking at 70 $^{\circ}$C for 5 min and 1 hour for the layer followed by nanoprinting and without nanoprinting, respectively. 
         
         \paragraph{Perovskite nano-imprint:} Before imprinting, the mold is functionalized by a silanization process to avoid sticking with perovskite. The imprinting process is operated in ambient atmosphere. The nano-imprint is conducted with the use of a Rondol manual press at 3 kN and 100 $^{\circ}$C  for 15 minutes. After 15 minutes, the heat is turned off and the pressure is released when the temperature reaches 35 $^{\circ}$C. 
         
        \paragraph{Experimental setup:} Perovskite metasurfaces are characterized using a house-made setup that measures the angle-resolved reflectivity and photoluminescece. Excitation (white light source for reflectivity and laser for photoluminescence) is focused onto the sample via a microscope objective (NA=0.45) and the signal is collected via the same objective in a confocal geometry. The back-focal plane of the objective is imaged onto the entrance slit of a spectrometer and the is dispersed onto a CCD sensor at the output of the spectrometer. With this configuration, the dispersion is measured along the direction that is imposed by the spectrometer slit. 
        
        \paragraph{Analytical model:} The couplings between the three modes (1,0), (-1,1) and (-1,-1) are described by the FW Hamiltonian (see details in the Supplemental Material):
        \begin{equation}\label{eq:mainFW_3x3}
            H_{FW}(\vec{k})=\begin{pmatrix}
                \omega_1(\vec{k})&U_1&U_1 \\ U_1&\omega_2(\vec{k})&U_2 \\ U_1&U_2&\omega_3(\vec{k})
            \end{pmatrix}
            +i\begin{pmatrix}
            \gamma_1&\Gamma_{12}(\vec{k})&\Gamma_{13}(\vec{k}) \\ \Gamma_{12}(\vec{k})&\gamma_2&\Gamma_{23}(\vec{k}) \\ \Gamma_{13}(\vec{k})&\Gamma_{23}(\vec{k})&\gamma_2
            \end{pmatrix}
        \end{equation}
        In the Hermitian part (first term), $\omega_{j=1..3}(\vec{k})$ is the dispersion of uncoupled modes [white dashed line in Fig~\ref{fig1}(b,c)]; $U_1$ is the diffractive coupling strength between (1,0) and (-1,$\pm$1); $U_2$ is the diffractive coupling strength between (-1,1) and (-1,-1). In the non-Hermitian part (second term), $\gamma_1$ and $\gamma_2$ are the radiative losses of (1,0) and (-1,$\pm$1) respectively. The coupling terms $\Gamma_{12(\vec{k})}$, $\Gamma_{13(\vec{k})}$ and $\Gamma_{23(\vec{k})}$ correspond to the radiative couplings (i.e. losses exchange mechanism) between the uncoupled modes. The analytical expressions of $\omega_{j=1..3}(\vec{k})$ and $\Gamma_{ij}(\vec{k})$ are given in the supplemental material. The real and imaginary part of the complex eigenvalues of \eqref{eq:mainFW_3x3} correspond to the energy and losses of the hybrid modes respectively. Moreover, the eigenvectors of \eqref{eq:mainFW_3x3} provide the near/far field patterns and the polarization texture of the hybrid modes.

        \paragraph{Numerical simulationss:} The RCWA simulations have been performed with the S\textsuperscript{4} package provided by the Fan Group at the Stanford Electrical Engineering Department.\cite{Liu2012} The refractive index and extinction coefficient of perovskite, used in  RCWA simulations, is obtained from ellipsometry measurement. The ellipsometry experiment has been performed on perovskite layer that is pressed by a flat mold.

	\begin{acknowledgement} The authors thank Xavier Letartre and Pierre Viktorovitch for fruitful discussions. This work benefited from the facilities of the Nanolyon technological platform, member of the Renatech+ and CARAT networks for all nanofabrication processes.  Simulations were carried out using the facilities of the PMCS2I (Ecole Centrale de Lyon). This work is partly supported by the French National Research Agency (ANR) under the projects POPEYE (ANR-17-CE24-0020) and EMIPERO (ANR-18-CE24-0016). It is also supported by the Auvergne-Rh\^{o}ne-Alpes region in the framework of PAI2020 and the Vingroup Innovation Foundation (VINIF) annual research grant program under Project Code VINIF.2021.DA00169
    \end{acknowledgement} 

\begin{suppinfo}
In the Supplemental Material document, a full derivation of the FW model and the formation of FW-BIC provided. We also include additional results on the perovskite deposition and nanoimprint, as well as photoluminescence images in real and momentum space of of the structure.
\end{suppinfo}    

\bibliography{Ref}

\begin{center}
	\textbf{\large --- SUPPLEMENTAL MATERIAL ---}
\end{center}

\setcounter{equation}{0}
\setcounter{figure}{0}
\setcounter{table}{0}
\setcounter{page}{1}

\renewcommand{\theequation}{S\arabic{equation}}
\renewcommand{\thefigure}{S\arabic{figure}}
\renewcommand{\bibnumfmt}[1]{[S#1]}
\renewcommand{\vec}[1]{\boldsymbol{#1}}
\DeclarePairedDelimiter\bra{\langle}{\rvert}
\DeclarePairedDelimiter\ket{\lvert}{\rangle}
\DeclarePairedDelimiterX\braket[2]{\langle}{\rangle}{#1 \delimsize\vert #2}

\begin{figure}[htb!]
		\begin{center}
			\includegraphics[width=16cm]{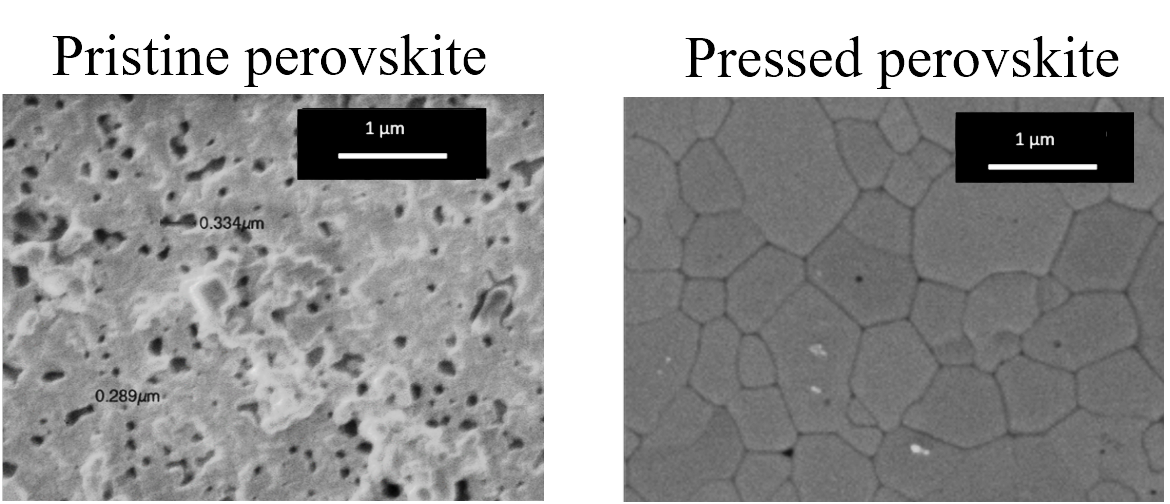}
		\end{center}
		\caption{\label{fig_S8}\textbf{Pristine vs Pressed perovskite layer}. SEM images of perovskite layer obtained by (left-panel): spincoating + anti-solvent dripping; (right-panel):  spincoating + anti-solvent dripping + pressing with flat mold under high pressure and high temperature.}
\end{figure}
\begin{figure}[htb!]
		\begin{center}
			\includegraphics[width=16cm]{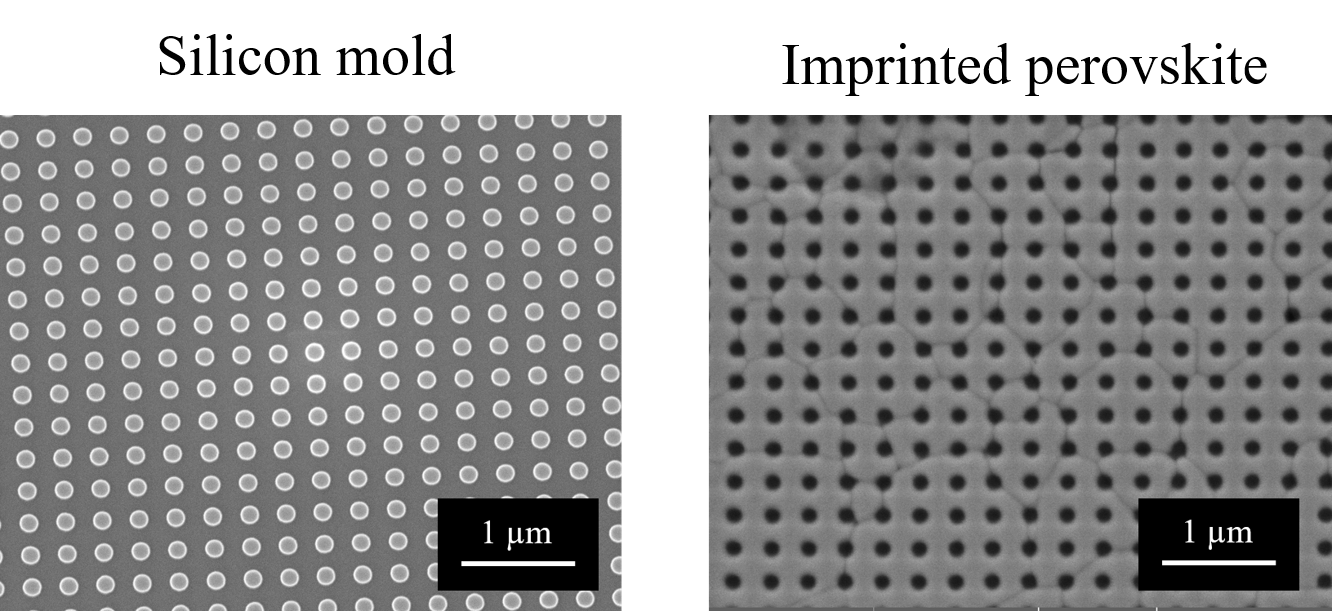}
		\end{center}
		\caption{\label{fig_S9}\textbf{Silicon mold vs Imprinted perovskite layer}. SEM images of the silicon mold and imprinted perovskite layer that is obtained by using the same mold.}
\end{figure}

\begin{figure}[htb!]
		\begin{center}
			\includegraphics[width=15cm]{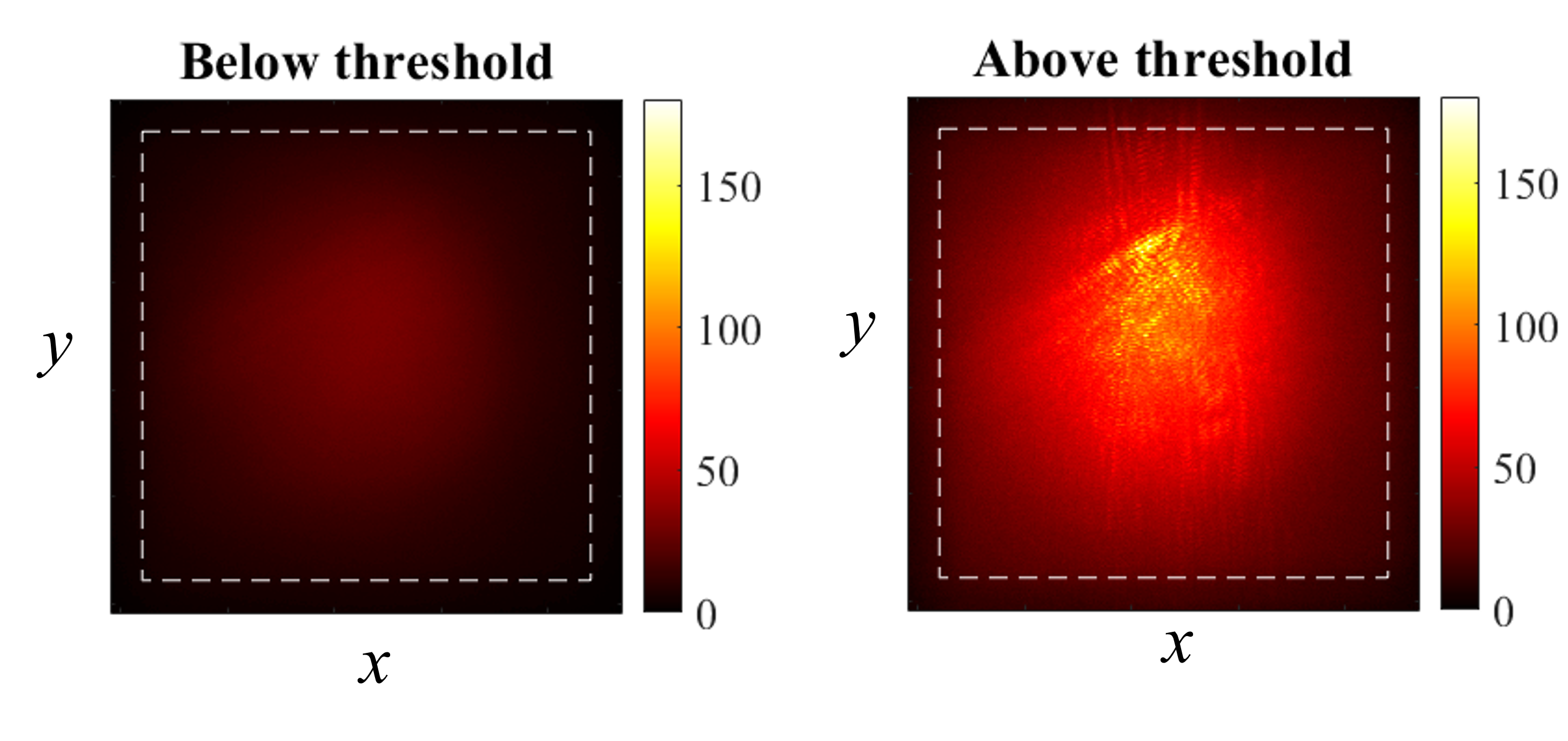}
		\end{center}
		\caption{\label{fig_S6}\textbf{Farfield imaging in real space}. Experimental image in real space of the photoluminescece signal (without any spectral filter and polarization elements) when pumping just below and just above threshold. The dashed line indicates the metasurface region where perovskite has been patterned.  We observe clearly a drastic enhancement of the photoluminesce signal and the appearance of speckles, indication of coherent emission, for the lasing emission.}
\end{figure}
\begin{figure}[htb!]
		\begin{center}
			\includegraphics[width=15cm]{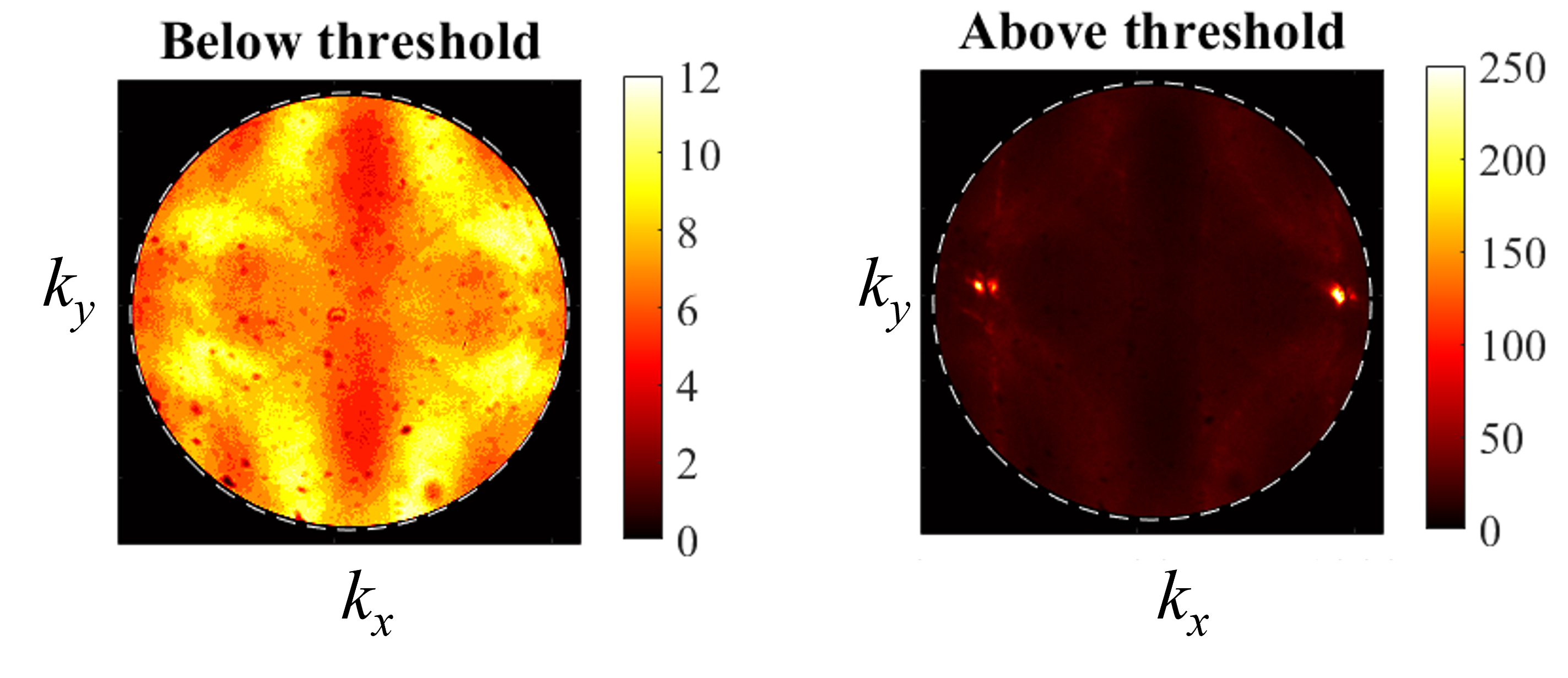}
		\end{center}
		\caption{\label{fig_S7}\textbf{Farfield imaging in momentum space}. Experimental image in momentum space of the photoluminescece signal when pumping below and above threshold. The dashed line indicates the numerical aperture of the microscope objective. The polarization is selected along y-axis. There is no spectral filter.  Below threshold, the photoluminesce takes place in a broad region of momentum space. Above threshold, we only observe a lasing emission that is concentrated in the vicinity of the twin FW-BICs.} 
\end{figure}

\section{Theoretical model for Friedrich-Wintgen couplings}
\label{sec:derivation}
In this section, we provide the detailed derivation of the effective Hamiltonian representing leaky Bloch modes in a general photonic lattice  of period $a_x = \eta\times a$ along x-axis, and $a_y=a$ along y-axis with $\eta\approx 1$. We limit our consideration for unit cells that exhibit lateral mirror symmetries $x\rightarrow -x$ and $y\rightarrow -y$.

In a perturbative approach, the Bloch resonances of the photonic lattice can be described in the basis of propagating in-plan guided modes in a dielectric slab of effective medium without corrugation. The Bloch resonances are results of the couplings within these guided modes and also between these guided modes with the radiative continuum. 

\subsection{Basis made of folded guides modes}
 For the sake of simplicity, we suppose that the dielectric slab of effective medium only exhibits a single guided mode in the spectral range of interest. The folding of this guided mode from the $(n,m)^{th}$ Brillouin zone given by the Bloch vector $\vec{B}_{n,m}=\frac{2\pi}{a_x}n \vec{u_x} + \frac{2\pi}{a_y} m\vec{u_y}$ to the first Brillouin zone correspond to the folded guided mode $\ket{n,m}$ of wavevector $\vec{k} = k_x \vec{u_x} + k_y \vec{u_x}$ with $k_{x,y} <  2\pi/a_{x,y}$.
 
 The ensemble of $\ket{n,m}$ with $n,m\in \mathbb{Z}$ is the basis of our theory. We note that for a general case of multi-mode operation, our theory can be easily extended by considering the basis made of $\ket{n,m}_p$ where $p$ is the index of the unfolded guided mode.
 
 Due to the folding, the propagation vector of the folded guided mode $\ket{n,m}$ is given by:
 \begin{equation}
 \vec{\beta}_{n,m}(\vec{k})=\frac{2\pi }{\eta a}n\vec{u_x} + \frac{2\pi}{a} m\vec{u_y} + \vec{k}  \label{eq:wavevector_origin}
 \end{equation} 
 
\subsubsection{Dispersion characteristic}
Assuming that the dispersion characteristic in the spectral range of interest of the unfolded guided mode is linear with the slope given by the effective group index $n_g$, one may show that the dispersion characteristic of $\ket{n,m}$ is given by:
 \begin{align}
 \begin{split}
     \omega_{n,m}(\vec{k})&=\omega_0 + \frac{c}{n_g}\left(\lvert \vec{\beta}_{n,m}(\vec{k}) \rvert-\frac{2\pi}{a}\right)\\
     &=\omega_0 +  \frac{2\pi c}{n_g a}\left[\sqrt{\left(\frac{n}{\eta} + \frac{k_x a}{2\pi}\right )^2 + \left(m+ \frac{k_y a}{2\pi}\right)^2} -1\right]  \label{eq:dispersion_origin}
     \end{split}
 \end{align}
 where $c$ is the speed of light and $\omega_0$ is the pulsation of the unfolded guided mode at $\lvert \vec{\beta}\rvert = 2\pi/a$.
 
To simplify the notations, it is preferable to work with dimensionless quantities. We thus redefine the wavevector and pulsation as:
\begin{align}
    \vec{q}&=\vec{k}a/2\pi
\end{align}
and
\begin{equation}
\hat{\omega}_{n,m}(\vec{q}) = \frac{\left[\omega_{n,m}(\vec{q})-\omega_0\right]n_g a}{2\pi c}. 
\end{equation}
Using these quantities, the propagation vector \eqref{eq:wavevector_origin} and the dispersion characteristic \eqref{eq:dispersion_origin} can be rewritten as:
 \begin{equation}
 \vec{\beta}_{n,m}(\vec{q})=\frac{2\pi}{a}\left[\left(\frac{n}{\eta} + q_x\right)\vec{u_x} + \left(m+ q_y\right)\vec{u_x} \right] \label{eq:wavevector_new}
 \end{equation} 
 and
\begin{equation}
 \hat{\omega}_{n,m}(\vec{q}) = \sqrt{\left(\frac{n}{\eta} + q_x\right )^2 + \left(m+ q_y\right)^2} -1 \label{eq:dispersion_new}  
\end{equation}
where $q_x=k_xa/2\pi$ and $q_y=k_ya/2\pi$.

From \eqref{eq:dispersion_new}, we deduce that the lowest bands in the folded dispersion band structure are $\ket{\pm 1,0}$, $\ket{0,\pm 1}$, $\ket{\pm 1, 1}$ and $\ket{\pm 1,-1}$.

 \subsubsection{Polarization vectors}
 Polarizations of guided modes are classified into TE and TM modes (i.e. the electric field $\vec{E}$ and magnetic field $\vec{H}$ are in-plan and perpendicular to the propagation vector respectively). Here we are only interested in the in-plane components of the electric field, given by in-plane polarization vector $\vec{u}_{n,m}^{TE}(\vec{q})$ and $\vec{u}_{n,m}^{TM}(\vec{q})$ for TE and TM guided modes respectively. Since $\vec{u}_{n,m}^{TE}(\vec{q}).\vec{\beta}_{n,m}(\vec{q})=0$, $\vec{u}_{n,m}^{TM}(\vec{q})||\vec{\beta}_{n,m}(\vec{q})$ and $\lvert \vec{u}_{n,m}^{TE}(\vec{q}) \rvert = \lvert \vec{u}_{n,m}^{TM}(\vec{q}) \rvert = 1$, from the propagation vector \eqref{eq:wavevector_new}, we obtain:
 \begin{equation}
    \vec{u}_{n,m}^{TE}(\vec{q})=\frac{1}{\sqrt{\left(\frac{n}{\eta} + q_x\right)^2 + \left(m+ q_y\right)^2}}
    \left(
    \begin{matrix}
    -(m+ q_y) \\
    \frac{n}{\eta} + q_x
    \end{matrix}
    \right)\label{eq:u_TE}
\end{equation} 
and
\begin{equation}
    \vec{u}_{n,m}^{TM}(\vec{q})=\frac{1}{\sqrt{\left(\frac{n}{\eta} + q_x\right)^2 + \left(m+ q_y\right)^2}}
    \left(
    \begin{matrix}
    \frac{n}{\eta} + q_x \\
    m+ q_y
    \end{matrix}
    \right). \label{eq:u_TM}
\end{equation}

In the following, we assume that the single unfolded mode is the fundamental TE mode, thus the TE and TM index will be omitted and the polarization vector $\vec{u}_{n,m}(\vec{q})$ is the one of TE modes $\vec{u}_{n,m}^{TE}(\vec{q})$.

\subsection{Coupling mechanism of folded guided modes}
The coupling between folded guided modes are possible if the phase matching conditions are fulfilled. There are two phase matching conditions:
\begin{itemize}
    \item \textit{Symmetry matching}: they are originated from unfolded guided modes of the same vertical symmetry parity if the structure exhibits a vertical symmetry mirror. Here we only considered folded guided modes from the same unfolded guided modes, thus this condition is automatically satisfied.
    \item \textit{Energy and Wavevector matchings}: the coupling is only effective in the vicinity of the crossing points of the dispersions.
\end{itemize}

\subsubsection{Diffractive coupling}
Thanks to the periodic corrugation, the folded guided modes can be coupled via near-field effect as soon as the phase matching conditions mentioned above are fulfilled. A diffractive coupling of strength $U_{ij}$ between two folded guided modes $\ket{i}=\ket{n_i,m_i}$ and $\ket{j}=\ket{n_j,m_j}$ would lead to an avoid crossing of band gap $\approx 2U_{ij}$ from a doubly degenerated crossing point between $\ket{i}$ and $\ket{j}$.

\subsubsection{Radiative losses and radiative couplings}
Due to the bandfolding, the dispersion of $\ket{n,m}$ can locate inside the light cone. Thus this folded guided mode can couple to the radiative continuum and exhibiting radiative losses of normalized rate $\gamma_{n,m}$ in the unit of $\frac{2\pi c}{n_g a}$.

A radiative coupling between two folded guided modes $\ket{i}=\ket{n_i,m_i}$ and $\ket{j}=\ket{n_j,m_j}$ is possible via the radiative continuum. This coupling, discovered by Friedrich and Wintgen in a general scheme, is the inteference between the two folded guided modes via the farfield radiation. One may describe the coupling strength $\Gamma_{ij}$ by a simple expression:
\begin{equation}
    \Gamma_{ij}=\sqrt{\gamma_i \gamma_j}.\cos\alpha_{ij} \label{eq:Gamma_ij}
\end{equation}
where $\gamma_i$ and  $\gamma_j$ are simplified notations of $\gamma_{n_i,m_i}$ and $\gamma_{n_j,m_j}$ respectively. The coefficient $\cos\alpha_{ij}=\vec{u}_{n_i,m_i}.\vec{u}_{n_j,m_j}$ is due to the fact that only the same electric field component can interfere. 

\subsubsection{Friedrich-Wintgen Hamiltonian}
In the vicinity of a $N$ times degenerated crossing point $\vec{q_c}$ between $N$ folded guided modes $\ket{1}$, $\ket{2}$, ...,$\ket{N}$, all of these modes can be coupled one to each other via diffractive and radiative couplings. In the basis of uncoupled modes $B=\left\{\ket{1},\ket{2}, ...\ket{N}\right\}$, the system is depicted by an non-Hermitian Hamiltonian, given by:
\begin{equation}
    H_{FW}(\vec{q})=H_0(\vec{q}) +  U(\vec{q}) + i\Gamma(\vec{q}) \label{eq:FW_general}
\end{equation}
The construction of this Hamiltonian is explained as the following:
\begin{itemize}
    \item The first term $H_0$ is the Hamiltonian of uncoupled modes and only containing diagonal elements, given by $H_{0(i=j)}=\hat{\omega}_{i}(\vec{q})$. Here $\hat{\omega}_{i}(\vec{q})$ is the simplified notation of $\hat{\omega}_{n_i,m_i}(\vec{q})$ given by \eqref{eq:dispersion_new}.
    \item The second term $U$ corresponds to diffractive couplings and only containing non-diagonal elements, given by $U_{i\neq j} = U_{ij}$. Here $U_{ij}$ is the diffractive coupling strength between $\ket{i}$ and $\ket{j}$. Of course $U_{ij}=U_{ji}$.
    \item   The last term $\Gamma$ corresponds to the radiative losses and radiative coupling. Its diagonal elements are the radiative losses of each mode, given by $\Gamma_{i=j}=\gamma_i$. Its non-diagonal elements are given by $\Gamma_{i\neq j}=\sqrt{\gamma_i \gamma_j}\cos\alpha_{ij}$ as \eqref{eq:Gamma_ij}.  Of course $\Gamma_{ij}=\Gamma_{ji}$.
\end{itemize}

\subsubsection{Eigenvalues of $H_{FW}$: radiative losses and FW conditions}

The diagonalization of $H_{FW}$ provides $N$ eigenmodes $\ket{\Omega_1}$, $\ket{\Omega_2}$... $\ket{\Omega_N}$ with corresponding eigenvalues $\Omega_1$, $\Omega_2$, ...,$\Omega_N$. These eigenvalues are complex numbers of which the real parts are pulsation  and the imaginary part are radiative losses. 

An eigenmode $\ket{\Omega_{BIC}}$ is a  Friedrich-Wintgen bound state in the continuum (FW-BIC) if the imaginary part of its eigenvalue is zero, i.e. $\text{Im}\left(\Omega_{BIC}\right)=0$. Such a destructive interference configuration requires special relations between the elements of $H_0$, $U$ and $\Gamma$. These relations are called Friedrich-Wintgen conditions (FW conditions). For example, in the case of a $2\times2$ $H_{FW}$, given by $H_{FW}=\begin{psmallmatrix}\hat{\omega}_1 & 0 \\ 0 & \hat{\omega}_2 \end{psmallmatrix} + \begin{psmallmatrix}0 & U \\ U & 0 \end{psmallmatrix} + i\begin{psmallmatrix}\gamma_1 & \sqrt{\gamma_1 \gamma_2}\cos\alpha \\ \sqrt{\gamma_1 \gamma_2}\cos\alpha & \gamma_2 \end{psmallmatrix}$, the FW conditions are $\sin\alpha=0$ and $\hat{\omega}_1-\hat{\omega}_2=(\gamma_1-\gamma_2)U/\sqrt{\gamma_1\gamma_2}\cos\alpha$.

\subsubsection{Eigenmodes of $H_{FW}$: eigenvectors, nearfield and farfield radiation}

The nearfield and farfield pattern of a given eigenmode $\ket{\Omega}$ can be calculated from the its eigenvector $\vec{v}_{\Omega}=\begin{psmallmatrix}C_1\\C_2\\...\\C_N\end{psmallmatrix}$ obtain from $H_{FW}$. Indeed, the decomposition of $\ket{\Omega}$ in the basis of folded guided modes is given by:
\begin{equation}
    \ket{\Omega}=\sum_{j=1}^N{C_j\ket{j}}
\end{equation}
Thus the nearfield and farfield pattern of $\ket{\Omega}$ is simply given by linear combination of the ones of the folded guided modes. 

The nearfield of the folded guided mode $\ket{j}$ is a plane wave propagating in-plane (x,y) with wavevector $\vec{\beta}_{n_j,m_j}(\vec{q})$  and vertical confinement profile $f(z)$. Thus it is given by:
\begin{align}
    \begin{split}
    \vec{\mathcal{E}}_{\ket{j}}^{\text{nearfield}}&=\mathcal{E}_0 f(z) \exp{\left[i\vec{\beta}_{n_j,m_j}(\vec{q}).\vec{r}_{||}\right]} \vec{u_{n_j,m_j}}\\
                            &=\mathcal{E}_0 f(z) \exp{\left(i\frac{2\pi \vec{q}.\vec{r}_{||}}{a}  \right)} \exp{\left(i\vec{B}_{n_j,m_j}.\vec{r}_{||}\right)} \vec{u_{n_j,m_j}}
\end{split}
\end{align}
where $\vec{r}_{||}=x\vec{u_x} + y\vec{u_y}$ is the in-plane coordinates; $\vec{B}_{n_j,m_j}=\frac{2\pi}{a_x}n_j \vec{u_x} + \frac{2\pi}{a_y} m_j\vec{u_y}$ is the Bloch vector of the band folding; $\vec{u_{n_j,m_j}}$ is the polarization vector from \eqref{eq:u_TE} for TE modes, and the constant $\mathcal{E}_0$ is a the field amplitude.

The Bragg scattering mechanism due to periodic corrugation folds the guided mode of propagation vector $\vec{\beta}_{n,m}(\vec{q})$ to the same wavevector $\frac{2\pi \vec{q}}{a}$ for any couples $(n,m)$. As a consequence, the folded guided mode $\ket{j}$ radiates to the free space as plane wave having $\frac{2\pi \vec{q}}{a}$ as the in-plane wavevector component. In the vicinity of the normal emission and of the crossing point, these radiating waves are approximately described as plane waves of the same wavevector $\frac{\omega_c}{c}\vec{u_z}$ while preserving the polarization $\vec{u}_{n_j,m_j}$ of the unfolded modes (here $\omega_c$ is the pulsation of the crossing point). Moreover, the amplitude of the radiating wave is dictated by the radiative losses, thus propotional to $\sqrt{\gamma_j}$. Therefore the  farfield of the folded guided mode $\ket{j}$ is given by
\begin{equation}
     \vec{\mathcal{E}}_{\ket{j}}^{\text{farfield}}\propto\mathcal{E}_0 \sqrt{\gamma_j} \exp\left(\frac{i\omega_cz}{c}\right) \vec{u}_{n_j,m_j}(\vec{q})
\end{equation}

Therefore,the nearfield and farfield pattern of the eigenmodes $\ket{\Omega}$ are given by:
\begin{align}
\vec{\mathcal{E}}_{\ket{\Omega}}^{\text{nearfield}}&=\mathcal{E}_0 f(z) \exp{\left(i\frac{2\pi \vec{q}.\vec{r}_{||}}{a}  \right)} \sum_{j=1}^{N}{C_j\exp{\left(i\vec{B}_{n_j,m_j}.\vec{r}_{||}\right)} \vec{u_{n_j,m_j}}} \label{eq:nearfield}\\
 \vec{\mathcal{E}}_{\ket{\Omega}}^{\text{farfield}}&\propto\mathcal{E}_0 \exp\left(\frac{i\omega_cz}{c}\right) \sum_{j=1}^{N}{C_j\sqrt{\gamma_j}\vec{u}_{n_j,m_j}(\vec{q})} \label{eq:farfield}
\end{align}
\begin{figure}[htb!]
		\begin{center}
			\includegraphics[width=8cm]{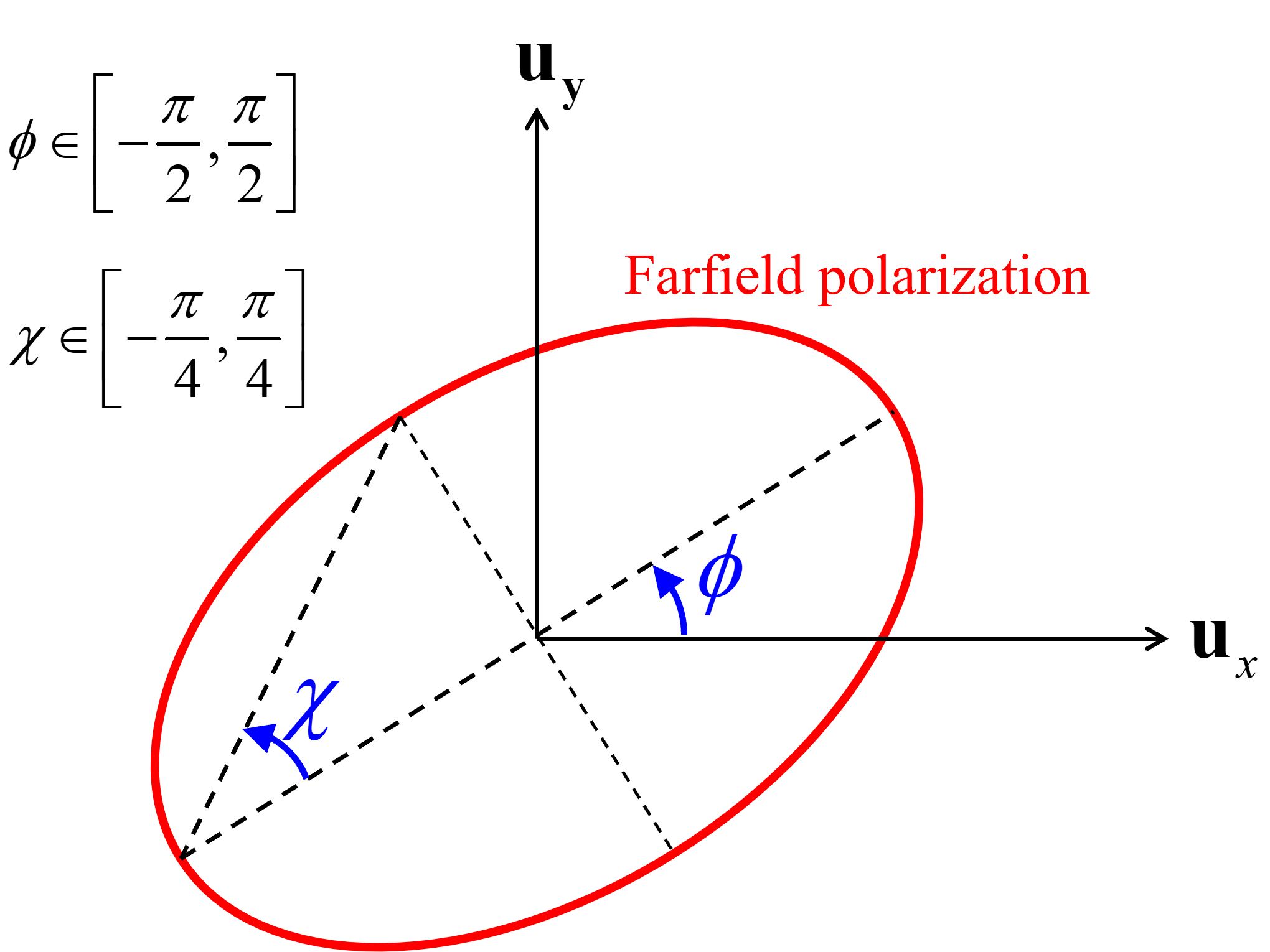}
		\end{center}
		\caption{\label{fig_S1}An arbitrary polarized light is determined by two angles $\phi$ and $\chi$ defining the orientation and ellipticity of the polarization.}
\end{figure}

From the distribution in momentum space of the complex electric field $\vec{\mathcal{E}}_{\ket{\Omega}}^{\text{farfield}}$, we can retrieve the its polarization texture given by the orientation angle $\phi$ and ellitpicity angle $\chi$ (see Fig.~\ref{fig_S1}):
\begin{align}
    \tan2\phi&=\frac{2\text{Re}(\mathcal{E}_{x}^*.\mathcal{E}_{y})}{|\mathcal{E}_{x}|^2-|\mathcal{E}_{y}|^2}\label{eq:phi}\\
    \sin2\chi&=\frac{2\text{Im}(\mathcal{E}_{x}^*.\mathcal{E}_{y})}{|\mathcal{E}_{x}|^2+|\mathcal{E}_{y}|^2} \label{eq:chi}
\end{align}
where $\mathcal{E}_{x}=\vec{\mathcal{E}}_{\ket{\Omega}}^{\text{farfield}}.\vec{u_x}$ and $\mathcal{E}_{y}=\vec{\mathcal{E}}_{\ket{\Omega}}^{\text{farfield}}.\vec{u_y}$.

Importantly, if the FW conditions are fulfilled and $\ket{\Omega}$ is a FW-BIC, one may show that $\sum_{j=1}^{N}{C_j\sqrt{\gamma_j}\vec{u}_{n_j,m_j}}=0$. Therefore the farfield radiation vanishes at FW-BIC. Moreover, if the FW-BICs are located at isolates position in the momentum space, they correspond to singularities of polarization vortex.

\subsection{Friedrich-Wintgen coupling between $\ket{1,0}$, $\ket{-1,1}$ and $\ket{-1,-1}$}
\subsubsection{Dispersion characteristics and polarization of uncoupled modes}
\begin{figure}[htb!]
		\begin{center}
			\includegraphics[width=12cm]{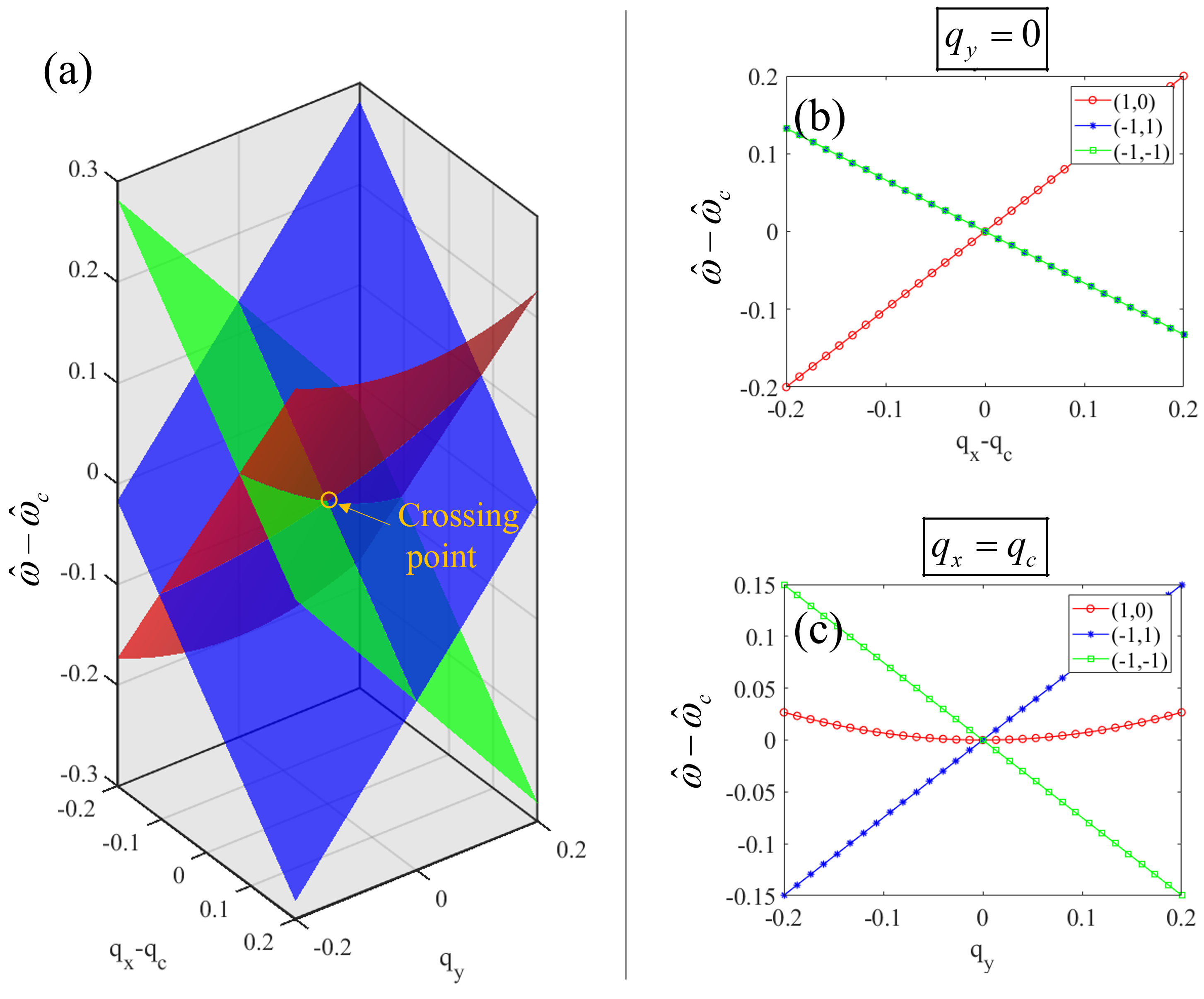}
		\end{center}
		\caption{\label{fig_S2}\textbf{Dispersion characteristics of the three folded guided modes $\ket{1}\equiv\ket{1,0}$, $\ket{2}\equiv\ket{-1,1}$ and $\ket{3}\equiv\ket{-1,-1}$}. (a) The dispersion surfaces in $(q_x,q_y)$ plane of . (b,c) Dispersion curves along $q_x$ with $q_y=0$. (c) Dispersion curves along $q_y$ with $q_x=q_c$}
\end{figure}

In previous sections, we have developped a general theory for Friedrich-Wintgen inteference between folded guided modes in rectangular lattice. In the following, we will focus on the coupling between the three TE folded guided modes given by:
\begin{align}
\begin{split}
    \ket{1}&\equiv\ket{n=1,m=0}\\
    \ket{2}&\equiv\ket{n=-1,m=1}\\
    \ket{3}&\equiv\ket{n=-1,m=-1}
\end{split}
\end{align}
From \eqref{eq:dispersion_new}, the dispersion of these modes are given by:
\begin{align}
    \begin{split}
 \hat{\omega}_1(\vec{q})&=\hat{\omega}_{1,0}(\vec{q})= \sqrt{\left(\frac{1}{\eta} + q_x\right )^2 + q_y^2} -1 \\
 \hat{\omega}_2(\vec{q})&=\hat{\omega}_{-1,1}(\vec{q})= \sqrt{\left(-\frac{1}{\eta} + q_x\right )^2 + \left(1+ q_y\right)^2} -1 \\   
  \hat{\omega}_3(\vec{q})&=\hat{\omega}_{-1,-1}(\vec{q})= \sqrt{\left(-\frac{1}{\eta} + q_x\right )^2 + \left(-1+ q_y\right)^2} -1 
 \end{split}
\end{align}
It can be easily demonstrated that the three mode degenerate triply at a crossing point of wavevector $\vec{q}_c=\begin{pmatrix}q_{c}=\eta/4\\0\end{pmatrix}$ and normalized pulsation $\hat{\omega}_c=\frac{1}{\eta} + \frac{\eta}{4} - 1$. In the vicinity of the triple crossing point, the dispersions are given by:
\begin{align}
    \begin{split}
    \hat{\omega}_1(\vec{q})&\approx \hat{\omega_c} + \left(q_x - q_{c}\right) + \frac{4+\eta^2}{8\eta}q_y^2\\
    \hat{\omega}_2(\vec{q})&\approx \hat{\omega_c} - \frac{4-\eta^2}{4+\eta^2}\left(q_x - q_{c}\right) + \frac{4\eta}{4+\eta^2}q_y\\
    \hat{\omega}_3(\vec{q})&\approx \hat{\omega_c} - \frac{4-\eta^2}{4+\eta^2}\left(q_x - q_{c}\right) - \frac{4\eta}{4+\eta^2}q_y
        \end{split}\label{eq:omega_123}
\end{align}
From these results, we deduce that in the vicinity of the crossing point, i) along $q_x$ with $q_y=0$: the three modes exhibit linear dispersion and the two modes $\ket{2}$ and $\ket{3}$ are degenerated; ii) along $q_y$ with $q_x=q_c$: $\ket{1}$ exhibits parabolic dispersion while $\ket{2}$ and $\ket{3}$ exhibit linear dispersions of opposite group velocity. The dispersion surfaces and different cuts along $q_x$ and $q_y$ are depicted in Fig.~\ref{fig_S2}.

From \eqref{eq:u_TE}, the polarization of these three modes are given by:
 \begin{align}
 \begin{split}
    \vec{u}_1(\vec{q})&=\frac{1}{\sqrt{\left(\frac{1}{\eta} + q_x\right)^2 +  q_y^2}}
    \left(
    \begin{matrix}
    -q_y \\
    \frac{1}{\eta} + q_x
    \end{matrix}
    \right)\\
    \vec{u}_{2,3}(\vec{q})&=\frac{1}{\sqrt{\left(\frac{1}{\eta} - q_x\right)^2 + \left(1\pm q_y\right)^2}}
    \left(
    \begin{matrix}
    \mp1 - q_y \\
    -\frac{1}{\eta} + q_x
    \end{matrix}
    \right)
    \end{split}\label{eq:u_123}
\end{align} 

\subsubsection{Friedrich-Wintgen Hamiltonian}
Due to the mirror symmetry $y\rightarrow -y$,  the diffractive coupling strength $U_{12}$ and $U_{13}$ are equal. For the same reason, $\ket{2}$ and $\ket{3}$ would have the same radiative losses (i.e. $\gamma_2=\gamma_3$). To simplify the notations, in the following we denote $U_1\equiv U_{13}=U_{12}$ and $U_2\equiv U_{23}$.

Therefore in our case, the Friedrich-Wintgen Hamiltonian from \eqref{eq:FW_general} is reduced to:
\begin{equation}\label{eq:FW_3x3}
H_{FW}(\vec{q})=\begin{pmatrix}
    \hat{\omega}_1(\vec{q})&U_1&U_1 \\ U_1&\hat{\omega}_2(\vec{q})&U_2 \\ U_1&U_2&\hat{\omega}_3(\vec{q})
\end{pmatrix}
+i\begin{pmatrix}
\gamma_1&\Gamma_{12}(\vec{q})&\Gamma_{13}(\vec{q}) \\ \Gamma_{12}(\vec{q})&\gamma_2&\Gamma_{23}(\vec{q}) \\ \Gamma_{13}(\vec{q})&\Gamma_{23}(\vec{q})&\gamma_2
\end{pmatrix}
\end{equation}
where
\begin{align}
\begin{split}
\Gamma_{12}(\vec{q})&=\vec{u}_1(\vec{q}).\vec{u}_2(\vec{q})\sqrt{\gamma_1\gamma_2}\\
\Gamma_{13}(\vec{q})&=\vec{u}_1(\vec{q}).\vec{u}_3(\vec{q})\sqrt{\gamma_1\gamma_2}\\
\Gamma_{23}(\vec{q})&=\vec{u}_2(\vec{q}).\vec{u}_3(\vec{q})\gamma_2
\end{split}
\end{align}

\subsubsection{Formation of Friedrich-Wintgen BIC}
Using \eqref{eq:omega_123} and \eqref{eq:u_123}, one may demonstrate that the FW conditions are established at $\vec{q}_{BIC}=\begin{pmatrix}q_{BIC}\\0\end{pmatrix}$ with $q_{BIC}\approx q_c$
; and the three corresponding complex eignevalues  given by:
\begin{align}
\Omega_1(\vec{q}_{BIC})&\approx \frac{U_1}{\sqrt{2}}\left(\sqrt{\frac{\gamma_2}{\gamma_1}}-\sqrt{\frac{\gamma_1}{\gamma_2}}\right) - U_1^2\left(\sqrt{\frac{\gamma_2}{\gamma_1}}+\sqrt{\frac{\gamma_1}{\gamma_2}}\right)^2 \label{eq:Omega1}\\
\Omega_2(\vec{q}_{BIC})&\approx \frac{U_1}{\sqrt{2}}\left(\sqrt{\frac{\gamma_2}{\gamma_1}}-\sqrt{\frac{\gamma_1}{\gamma_2}}\right) + U_1^2\left(\sqrt{\frac{\gamma_2}{\gamma_1}}+\sqrt{\frac{\gamma_1}{\gamma_2}}\right)^2 + i(\gamma_1 + \gamma_2) \label{eq:Omega2}\\
\Omega_3(\vec{q}_{BIC})&\approx \sqrt{2}U_1\left(\sqrt{\frac{\gamma_2}{\gamma_1}}-\sqrt{\frac{\gamma_1}{\gamma_2}}\right) - 2U_2 + i\gamma_2 \label{eq:Omega3}
\end{align}

\begin{figure}[htb!]
		\begin{center}
			\includegraphics[width=16cm]{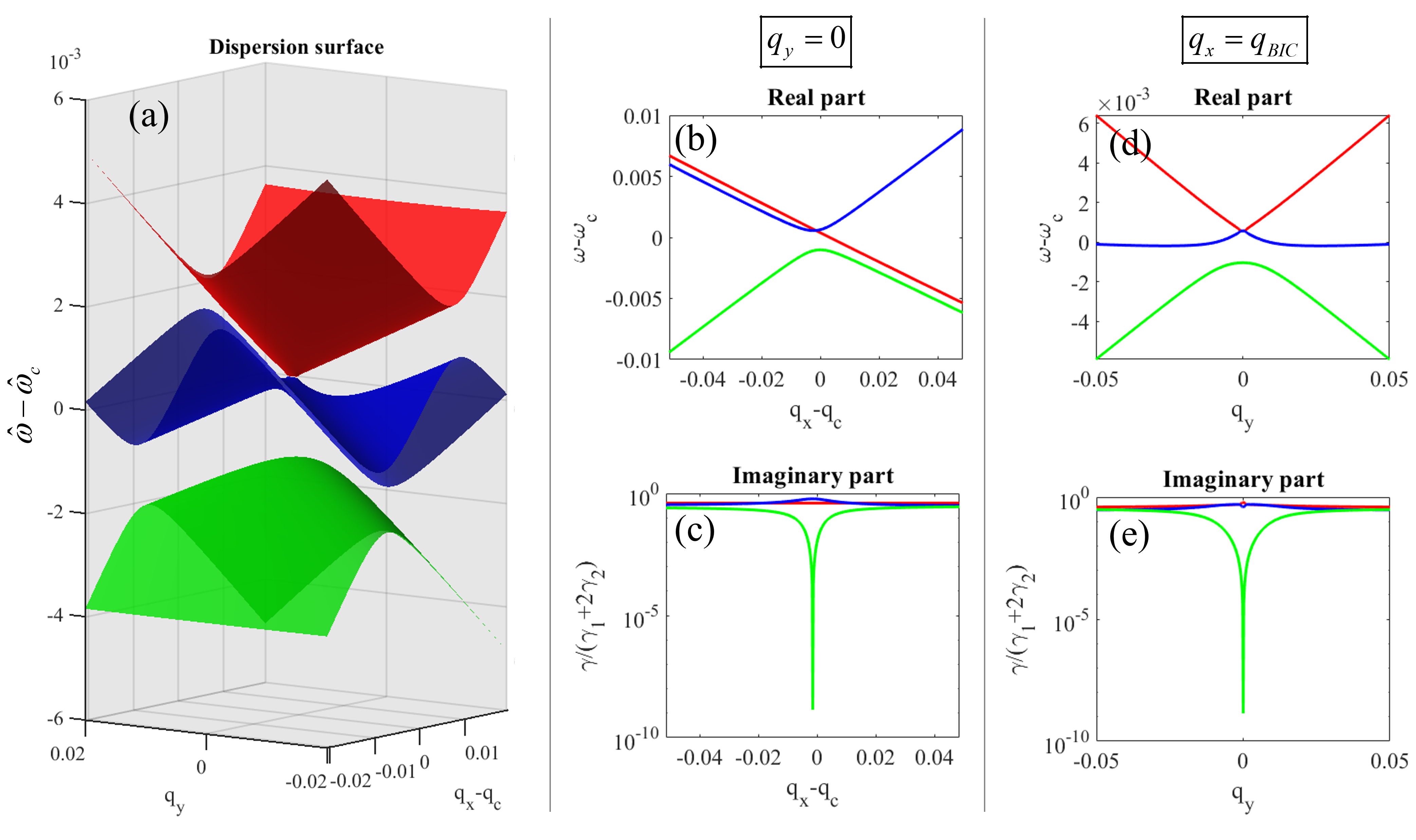}
		\end{center}
		\caption{\label{fig_S3}\textbf{Dispersion characteristics of eigenmodes $\ket{\Omega_1}$, $\ket{\Omega_2}$ and $\ket{\Omega_3}$ resulted from the FW Hamiltonian \eqref{eq:FW_3x3}}. (a) The dispersion surfaces of $\text{Re}(\Omega_{1,2,3})$ in $(q_x,q_y)$ plane. (b,c) Dispersion curves along $q_x$ with $q_y=0$ of  (b) $\text{Re}(\Omega_{1,2,3})$ and (c) $\text{Im}(\Omega_{1,2,3})$. (d,e) Dispersion curves along $q_y$ with $q_x=q_c$ of (d) $\text{Re}(\Omega_{1,2,3})$ and (e) $\text{Im}(\Omega_{1,2,3})$. The calculation has been done with $\eta=0.9$, $U_1=5.7\times10^{-4}$, $U_2=U_1/1.5$, $\gamma_1=0.4\lvert U_1\rvert$ and $\gamma_2=0.5\lvert U_1\rvert$.}
\end{figure}

Here the BIC state corresponds to $\ket{\Omega_1(\vec{q}_{BIC})}$ that exhibits zero losses. Figure~\ref{fig_S3} represent the dispersion characteristics of the eignmodes $\ket{\Omega_1}$, $\ket{\Omega_2}$ and $\ket{\Omega_3}$ of the FW Hamiltonian. The formation of FW-BIC in the vicinity of the crossing point is clearly evidenced.

\begin{figure}[htb!]
		\begin{center}
			\includegraphics[width=16cm]{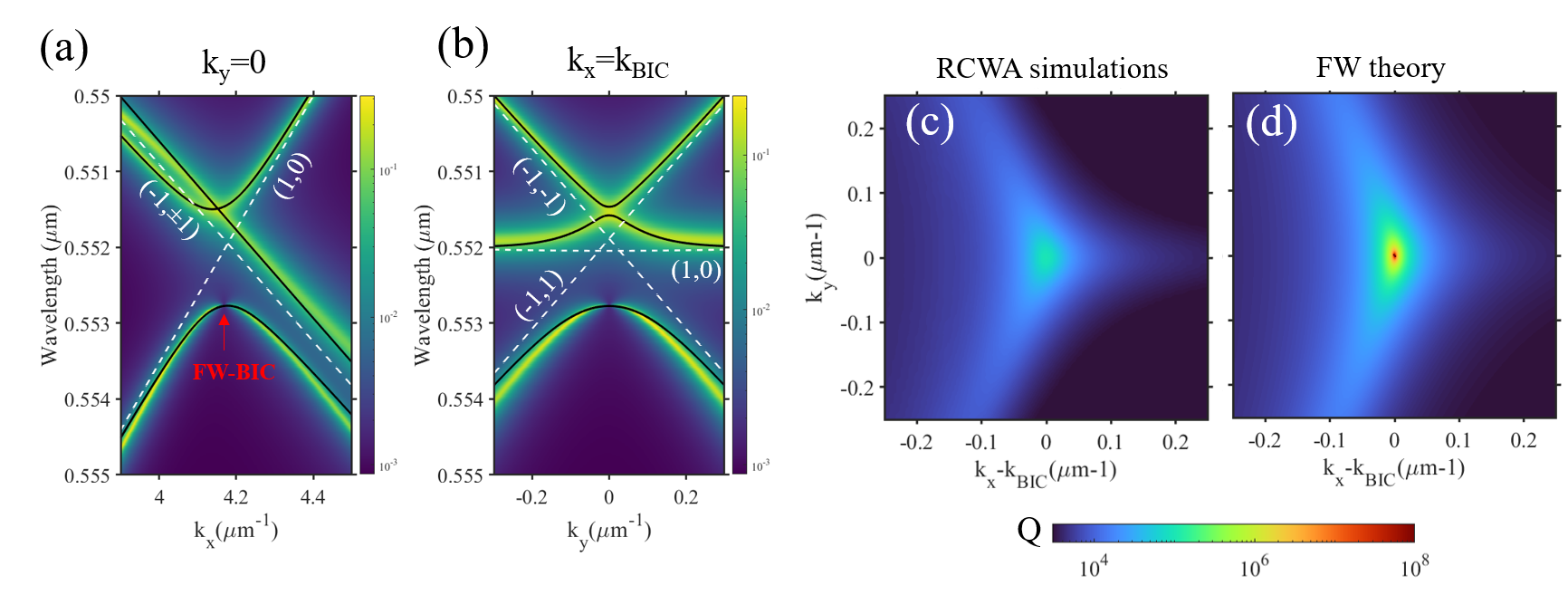}
		\end{center}
		\caption{\label{fig_S4}\textbf{Dispersion characteristics: RCWA vs FW model}. (a,b) The dispersions  along (a) $k_x$ and (b) $k_y$ obtained from RCWA simulations (pseudo colormap of resonances) and the real part of the eigenmodes obtained from the FW theory (black solid lines). Uncoupled folded modes are also presented by white dashed lines. (c,d) Quality factor mappings of the eigenmode $\ket{\Omega_1}$ that is extracted from (a) RCWA simulations and (b) FW theory.  The FW model is calculated with $n_g=5.5$, $\eta=0.9$, $U_1=5.7\times10^{-4}$, $U_2=U_1/1.5$, $\gamma_1=0.4\lvert U_1\rvert$ and $\gamma_2=0.5\lvert U_1\rvert$.}
\end{figure}

We now confront the results of our theoretical model with the ones obtained from numerical simulations using RCWA methods. Figures~\ref{fig_S4}(a,b) represent the dispersion along $k_x$ and $k_y$ obtained from RCWA simulations of angle-resolved abosrption (pseudo colormap of resonances) and the real part of the eigenmodes obtained from the FW theory (black solid lines), as well as the uncoupled folded modes (white dashed lines). It shows that the resonant energies are perfectly reproduced by the FW theory. Figures~\ref{fig_S4}(c) and (d) represent the quality factor mappings of $\ket{\Omega_1}$ obtained from  RCWA simulations and the FW theory respectively. The two mappings both exhibit maximum of quality factor at $\vec{k}_{BIC}$ and are in good agreement for the variation of quality factor when going out from $\vec{k}_{BIC}$. However,  while the quality factor from FW theory diverges at $\vec{k}_{BIC}$, the one from RCWA simulation is bounded at $10^5$. Therefore, the FW-BIC in becomes a quasi-BIC for the real structure. The residual losses, although very small, is due to parasitic absorption
of the perovskite material bellow its band-gap and the coupling with other modes that are not considered in the $3\times3$ model. 

We note that a twin FW-BIC is expected at $-\vec{k}_{BIC}$ due to the mirror symmetry $x\rightarrow -x$. This twin FW-BIC is resulted from the coupling between $\ket{\tilde{1}}=\ket{-1,0}$, $\ket{\tilde{2}}=\ket{1,1}$ and $\ket{\tilde{3}}=\ket{1,-1}$.

\subsubsection{Polarization texture of the FW-BIC twins}
\begin{figure}[htb!]
		\begin{center}
			\includegraphics[width=16cm]{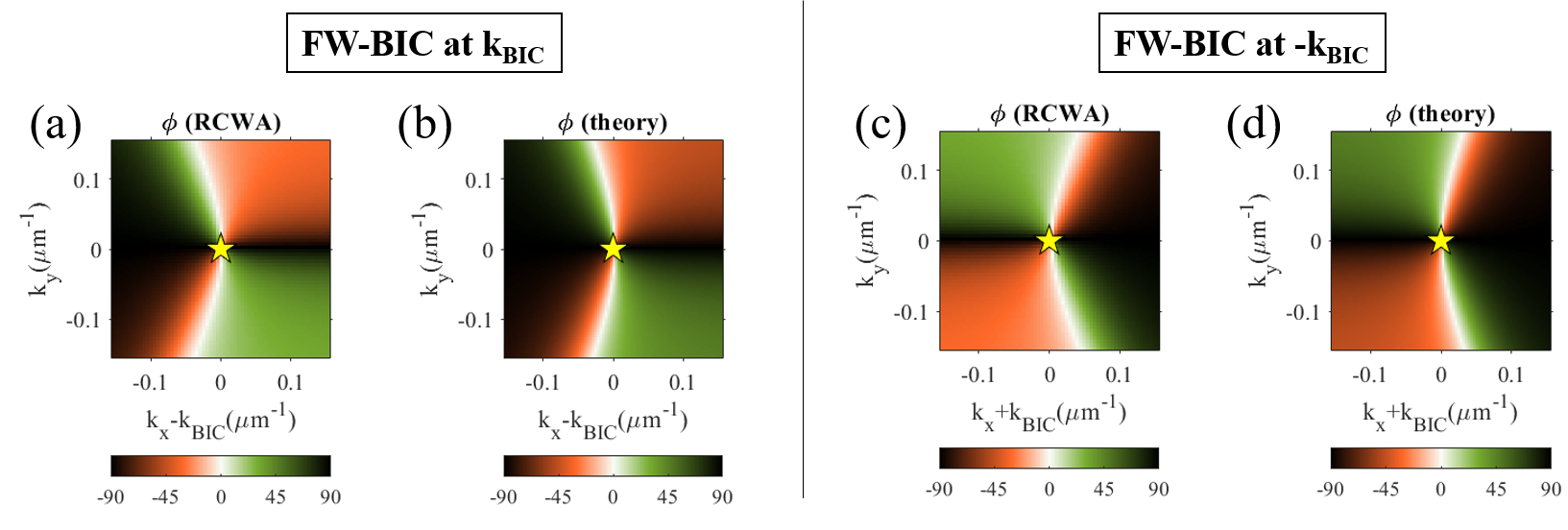}
		\end{center}
		\caption{\label{fig_S5}\textbf{Polarization textures: RCWA vs FW model}. Mapping of the polarization orientation angle $\phi$ in the vicinity of the FW-BIC located at (a,b)$\vec{k}_{BIC}$, (c,d)$-\vec{k}_{BIC}$. These textures are obtained from (a,c) RCWA simulations, (b,d) the FW theory. The parameters for the FW theory are the same as the one used in Fig.~\ref{fig_S3}.}
\end{figure}

Using \eqref{eq:farfield}, one can obtain the farfield patterns from the eigenvectors of \eqref{eq:FW_3x3}. Then the polarization textures can be calculated using \eqref{eq:phi}. Figures~\ref{fig_S5}(a-d) depict the polarization texture, obtained by RCWA simulations and FW theory, in the vicinity of the two FW-BICs. The results show a perfect agreement between the numerical simulations and the FW model. Most importantly, we evidence polarization singularity pinned at the FW-BIC location.

\end{document}